\documentclass[a4paper,11pt]{article}
\usepackage{jinstpub} 
\usepackage{todonotes}

\newcommand{\cevns}{CE$\nu$NS}

\title{\boldmath Calibration and characterization of the RED-100 detector at the Kalinin Nuclear Power Plant}

\author[a]{D.Yu.~Akimov,}
\author[a,c]{I.S. Aleksandrov,}
\author[a]{F.B.~Ata-Kurbonova,}
\author[b,a]{V.A.~Belov,}
\author[a]{A.I.~Bolozdynya,}
\author[b,a]{A.V.~Etenko,}
\author[a]{A.V.~Galavanov,}
\author[a,d]{Yu.V.~Gusakov,}
\author[a]{A.V.~Khromov,}
\author[f,a]{A.M.~Konovalov,}
\author[e,a]{V.N.~Kornoukhov,}
\author[b,a]{A.G.~Kovalenko,}
\author[a]{E.S.~Kozlova,}
\author[a]{Yu.I.~Koskin,}
\author[a]{A.V.~Kumpan,}
\author[a,g]{A.V.~Lukyashin,}
\author[a]{A.V.~Pinchuk,}
\author[a,1]{O.E.~Razuvaeva,\note{Corresponding author.}}
\author[a,\dag]{D.G.~Rudik, \note[\dag]{Now at: University of Naples Federico II, Corso Umberto I 40, Naples, 80138, Italy}}
\author[a]{A.V.~Shakirov,}
\author[b,a]{G.E.~Simakov,}
\author[a]{V.V.~Sosnovtsev,}
\author[a]{and A.A.~Vasin}

\affiliation[a]{National Research Nuclear University MEPhI (Moscow Engineering Physics Institute), \\ 31 Kashirskoe hwy, Moscow 115409, Russia}

\affiliation[b]{National Research Center “Kurchatov Institute”, \\
1 Akademika Kurchatova sq, Moscow, 123182, Russia}

\affiliation[c]{National Research Tomsk Polytechnic University,\\
30 Lenina ave, Tomsk, 634050, Russia}

\affiliation[d]{Joint Institute for Nuclear Research,\\ 6 Joliot-Curie st, Dubna, Moscow region 141980, Russia}

\affiliation[e]{Institute for Nuclear Research, \\7a 60-letiya Oktyabrya ave, Moscow, 117312, Russia}

\affiliation[f]{Lebedev Physical Institute, Moscow, 119991, Russian Federation, \\ 53 Leninskiy ave, Moscow, 119991, Russia}

\affiliation[g]{MIREA---Russian Technological University, Lomonosov Institute of Fine Chemical Technologies\\ 86 Vernadsky ave, Moscow, 119571, Russia}
\emailAdd{OERazuvaeva@mephi.ru}

 \abstract{RED-100 is a two-phase Xe detector designed and built for the study of coherent elastic neutrino-nucleus scattering (\cevns{}) of reactor antineutrinos. A comprehensive calibration was performed in order to obtain important parameters of the detector during its operation at the Kalinin Nuclear Power Plant (Tver region, Russia). This paper outlines the analysis of calibration data, position and energy reconstruction procedures, and calculation of the efficiency of electron extraction from the liquid xenon to the gas phase.}


\arxivnumber{2403.12645} 

\begin{document}
\maketitle
\flushbottom

\section{Introduction}
\label{sec:intro}

RED-100 is a two-phase emission detector designed and constructed for studying the coherent elastic neutrino-nucleus scattering (\cevns{}) process~\cite{Freedman, Kopeliovich:1974mv} with a xenon target~\cite{Akimov2017}. \cevns{} was observed for the first time in 2017 by the COHERENT collaboration~\cite{COHERENT:2017ipa}. At the present moment, many experiments are carried out with nuclear reactors as neutrino sources~\cite{Belov_2015, Aguilar-Arevalo_2016, ricochet, Buck_2020, Singh:2016glu, Strauss_2020, Chaudhuri:2022pqk}.
The detection of \cevns{} signal is quite complicated due to the very low energy (1--10~keV) of nuclear recoils and the technical difficulties in detecting signals of such low energy with massive detectors.
The RED-100 detector benefits from the high \cevns{} cross-section on a heavy xenon nucleus as well as a low energy threshold and large active mass of two-phase detectors ~\cite{Dolgoshein, 467913}. The use of this technique in the direct dark matter search resulted in significant progress~\cite{DarkSide:2022dhx,PandaX-4T:2021bab,XENON:2023cxc,LZ:2022lsv} making it reasonable to apply a similar approach to the challenge of \cevns{} detection.

The prediction of the~\cevns{} signal in units of ionization electrons depends on the parameters of detector operation, such as free electron lifetime in the liquid Xe till capture by electronegative impurities and efficiency of electron extraction to electroluminescence gas gap. Further sensitivity analysis, which will be described in a separate paper, includes a comparison of the background and simulated~\cevns{} spectra. To perform reliable simulation, detailed information about light generation and collection is required. It is contained in signal parameters for an electroluminescence from a single ionization electron and light collection parameters which are measured and evaluated with corresponding calibration data. Despite the fact that the expected \cevns{} signal has very low energy, the calibration in the MeV region is quite important for the light collection and electron extraction efficiencies calculations since these values are independent of energy and thus we can use a more convenient energy range for calibration. 

The RED-100 experiment is conducted at the Kalinin Nuclear Power Plant (KNPP, Tver region, Russia)~\cite{The_RED100_Experiment}. The detector is located at a distance of 19 meters under the reactor core. A comprehensive characterization of the detector response is performed \textit{in situ} with the goals of energy calibration and evaluation of the detector parameters crucial for the detailed simulation of the \cevns{} signal. This work outlines several types of calibrations performed for this purpose.

\section{The RED-100 detector}
\label{sec:calibr}
The active medium of the RED-100 detector is 130 kg of liquid xenon (LXe) contained in a cylindrical volume with a 12-sided Teflon reflector \cite{Akimov2017}. Electric fields in the drift volume and the electroluminescence gap are provided by several grid electrodes and field-shaping rings. The scintillation and electroluminescent signals are detected by two 19-unit arrays of 3" HAMAMATSU R11410-20 photomultipliers (PMTs) \cite{Akimov:2016eii, Akimov:2017coe}. In the bottom array, only seven central PMTs are used: four with nominal voltage for a data acquisition (DAQ) trigger by a prompt scintillation signal and the other three with reduced voltage participate in the muon veto. Also, there is an external fast amplifier, Phillips Scientific 777, set to 10x gain for each PMT channel. More information about the PMTs characterization can be found in ref.~\cite{Akimov:2016eii}.

\begin{figure}[htbp]
  \includegraphics[width=0.49\linewidth]{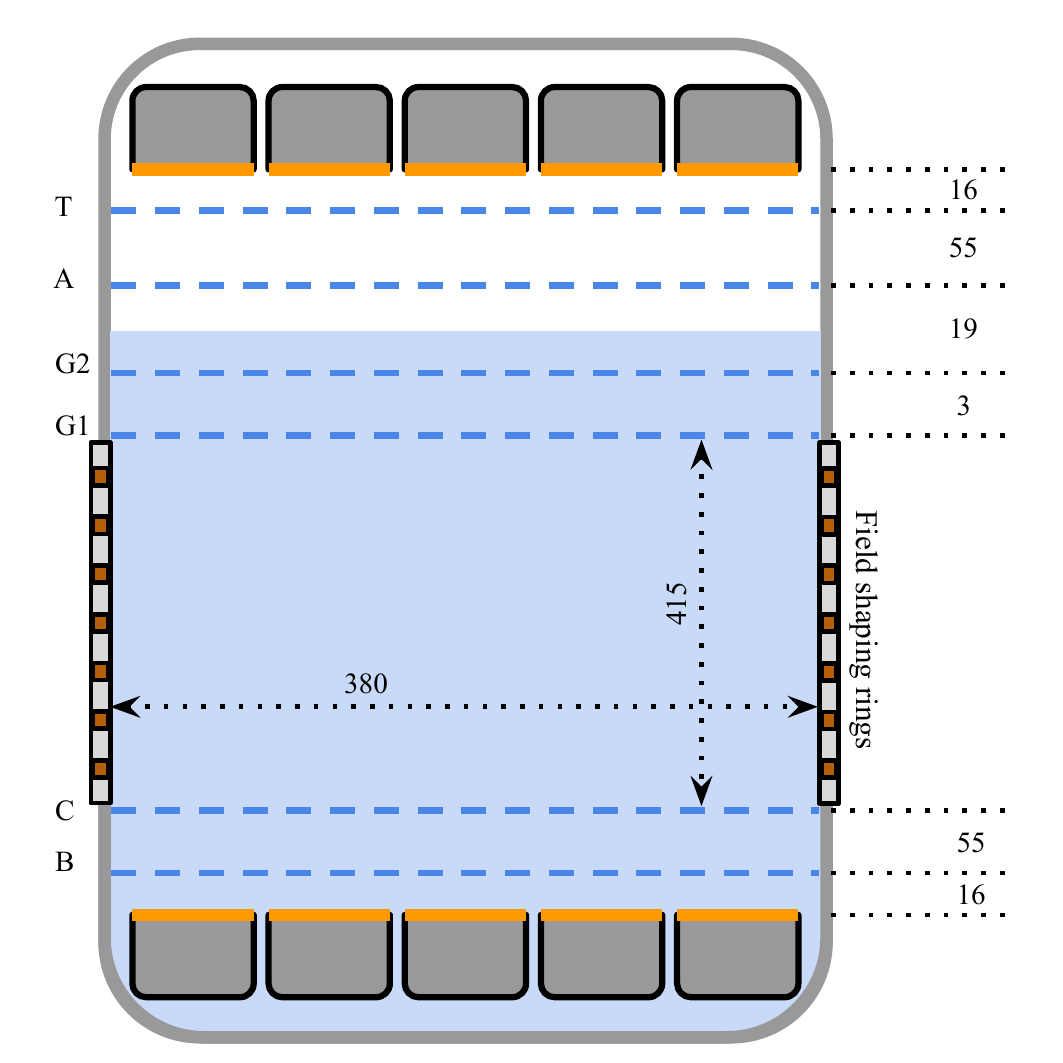}
  \hfill
  \includegraphics[width=0.46\linewidth]{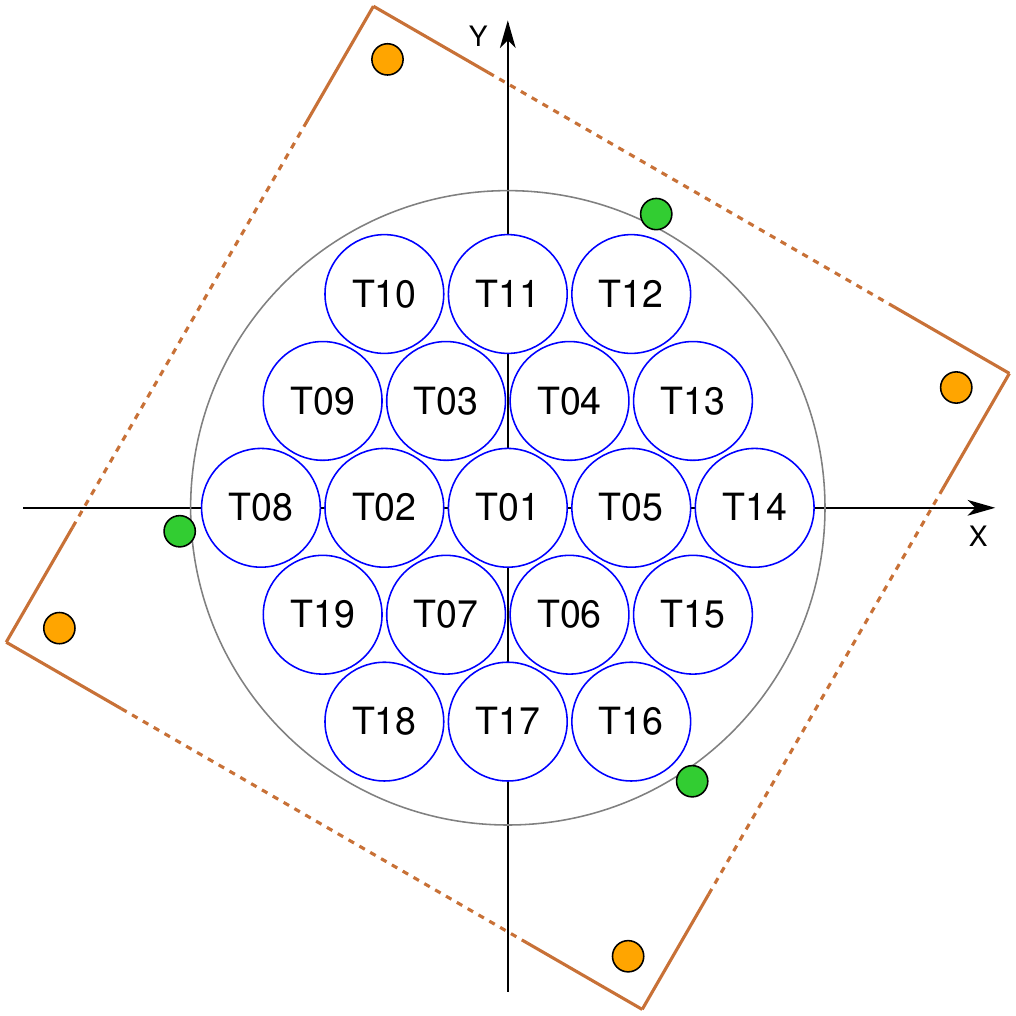}
  \caption{\textbf{Left}: Geometry of the RED-100 target volume and electrodes. A---anode; C---cathode; B,T---shielding electrodes; G1---shutter electrode, G2---extraction grid. Dimensions are given in millimeters. \\ \textbf{Right}: Scheme of PMT layout in the top array. Green circles indicate LEDs location. Brown lines indicate the edges of the copper passive shield. The dashed part of these lines indicates that the shielding is drawn out of scale. The length of the inner side of the square is 75 cm. Orange circles indicate source tubes positions.}
  \label{img:red100geometry}  
\end{figure}

The geometry of the target volume and the relative disposition of grids and PMTs are shown in figure \ref{img:red100geometry} (not to scale). 
In addition to the typical set of grids, there is a shutter electrode (G1). It was added to temporarily block the drift of ionization electrons to the surface of the liquid, which is important for the suppression of the spontaneous SE emission after a muon passage through the detector. This technology was patented by the RED-100 collaboration and described in ref.~\cite{RED100_2019}.
The detector operates with a drift field of 218 V/cm. Special care must be taken to the calculation of the electric field at the surface since it is close enough to the gate electrode to be affected by the non-uniformity of the field arising from the grid cells. We use the iterative algorithm described in ref.~\cite{READ1999363} to account for this effect. The calculated value of the extraction electric field is $2.68\pm0.04$~kV/cm for the current study and $2.82\pm0.04$~kV/cm for the previous measurements in the laboratory in 2019~\cite{RED100_2019}. The uncertainty of the obtained electric field value is related to the inaccuracy of the electroluminescence gap measurement and the dielectric constant of LXe~\cite{10.1063/1.1724850,doi:10.1139/p70-033}.

The detector is surrounded by a passive shield, which includes 70 cm of water and 5 cm of copper, to reduce the external background~\cite{shielding}. The reactor core and the power unit building serve as an additional shield (approximately 50 m.w.e. in the vertical direction as measured by the DANSS experiment~\cite{Alekseev_2016}) from atmospheric muons and neutrons generated by cosmic rays. Background conditions at the site were carefully measured and are described in~\cite{Akimov_2023}.

The energy deposited in the interaction of an ionizing particle with a target medium goes mostly to scintillation (S1) and ionization. The S1 signal is caused by the deexcitation of excited states of formed Xe dimers and occurs shortly after the moment of interaction. Part of the ionization electrons recombines and contributes to the S1 signal by forming Xe dimers, whereas the electrons that escape recombination drift towards the surface of the liquid under the applied electric field.
During the drift, they can be captured by electronegative impurities, leading to a characteristic exponential decrease of the signal with the depth of an interaction. At the surface, the electrons are extracted to the gas phase due to the higher electric field generated by the "gate" grid and the anode, or they can remain trapped under the surface \cite{Akimov:2012zz}. Extracted electrons produce a delayed electroluminescence signal (S2) in the gap between the surface and the anode grid~\cite{Aprile:2008bga}. The electrons that are originally trapped or captured by electronegative impurities can later escape and contribute to the specific background in the form of spontaneous emission of single electrons \cite{Akimov_2016_delayed_electrons, PhysRevD.102.092004, Kopec_2021}.

\cevns{} events result in very low energy depositions. In this region, S2 is a result of electroluminescence from several individual ionization electrons. This signal appears as several dozen of single photo-electrons (SPEs) scattered over a noticeable time span and across many channels.
To effectively handle this situation, fast electronics are used, and detailed waveforms with a sampling period of 2~ns are recorded for all PMTs by the data acquisition system (DAQ) for further processing and analysis \cite{Akimov2017}. The recorded time window varies for different types of measurements.
The trigger is either external from the pulse generator or internal from the leading-edge discriminator, tuned to start from S1 and run on a sum of the bottom PMTs.

As stated in the introduction, the simulation of the expected signal and data analysis require a precise understanding of the processes in the detector. For this purpose, the following quantities have to be obtained:
\begin{itemize}
    \item Single photo-electron (SPE) signal area for each PMT,
    \item Electron lifetime,
    \item Light response functions of PMTs,
    \item Parameters of electroluminescence from a single electron (SE)
    \item Efficiency of electron extraction from liquid to the gas gap (EEE).
\end{itemize} 

These quantities are measured using four different types of calibrations:
\begin{enumerate}
    \item \textbf{LED (light-emitting diode) calibration.} Three LEDs are located in the detector as shown in figure~\ref{img:red100geometry}. The LEDs are powered by low-intensity wide pulses from the generator to produce separate SPE signals in all channels. The LED pulses and the trigger are synchronized and are running at 2 Hz. This calibration data is used to measure SPE pulses and calculate SPE parameters for each PMT.

     \item \textbf{Muon calibration.} The detector is located at a relatively low overburden with a significant cosmic muon flux. An average muon signal can be used for the electron lifetime calculation. The muon calibration is performed with low voltage on the PMTs.

    \item \textbf{SE calibration.} The SE data are collected using an external trigger from the 2 Hz pulse generator. The main objective of this data is to facilitate the observation of accidental SPE signals and spontaneous SE events without any hardware threshold, which is very important for the SE parameters evaluation. 

    \item \textbf{Gamma calibration.} Radioactive sources of $^{60}$Co and $^{137}$Cs are placed inside the passive shielding via the guide tubes as illustrated in figure \ref{img:red100geometry}. The sources are aligned with the center of the detector's sensitive volume in a vertical direction. This calibration is performed once per week during the entire RED-100 data-taking period. The main objectives of this calibration include evaluation of EEE and the light response functions (LRFs) of the PMTs.

\end{enumerate}

\section{Data processing and analysis}
\label{sec:processing}
\subsection{REDOffline}
\label{sec:RED offline}
The dedicated software package REDOffline is developed by the collaboration for data processing. 
This lightweight and flexible solution features a modular design that benefits from the use of ROOT libraries \cite{BRUN199781} for various tasks. 
REDOffline converts raw waveform data into physical information about detected signals through a series of steps. Initially, the baseline is corrected to remove the remaining DC offset and low (18 kHz) frequency pickup noise. The correction is achieved by constructing a spline approximation of the pickup and subsequently subtracting it. 
The spline knots are placed every 3.75 $\mu$s, which is wider than the measured S2 width to ensure that individual SE or S2 signals do not violate the procedure.
The influence of the pickup on the trigger is negligible due to the typical signal size in the muon and gamma calibration data. In the LED and SE calibration, the trigger is external from the pulse generator.

A simple over-threshold software trigger is applied for pulse detection. 
The threshold is individual for each channel and is based on the measured noise level in that channel. The typical threshold value is about 1.8~mV. 
The pulse time region is set slightly wider than the threshold range to account for pulse samples below the threshold. 
Additionally, there is a limitation that the pulse duration must be at least 40 ns. 
For each pulse, multiple parameters are evaluated, including amplitude, area, start time, and duration. 
The pulse area is calculated as a simple sum of waveform samples within the pulse time region multiplied by the sampling period. 
The effective threshold for pulse finding is low enough to detect SPEs with an efficiency of about 97\%.

The subsequent clustering procedure groups pulses related to the same physical signal---significantly large S1 and S2 signals from $\gamma$-events and groups of SPE pulses from SE---in one or many channels. 
The algorithm varies depending on the type of signal and is described in the corresponding sections below.

\subsection{SPE (single photo-electron) parameters}
\label{subsec_SPEparams}
\begin{figure}[htbp]
 \centering
 \includegraphics[width=0.49\linewidth]{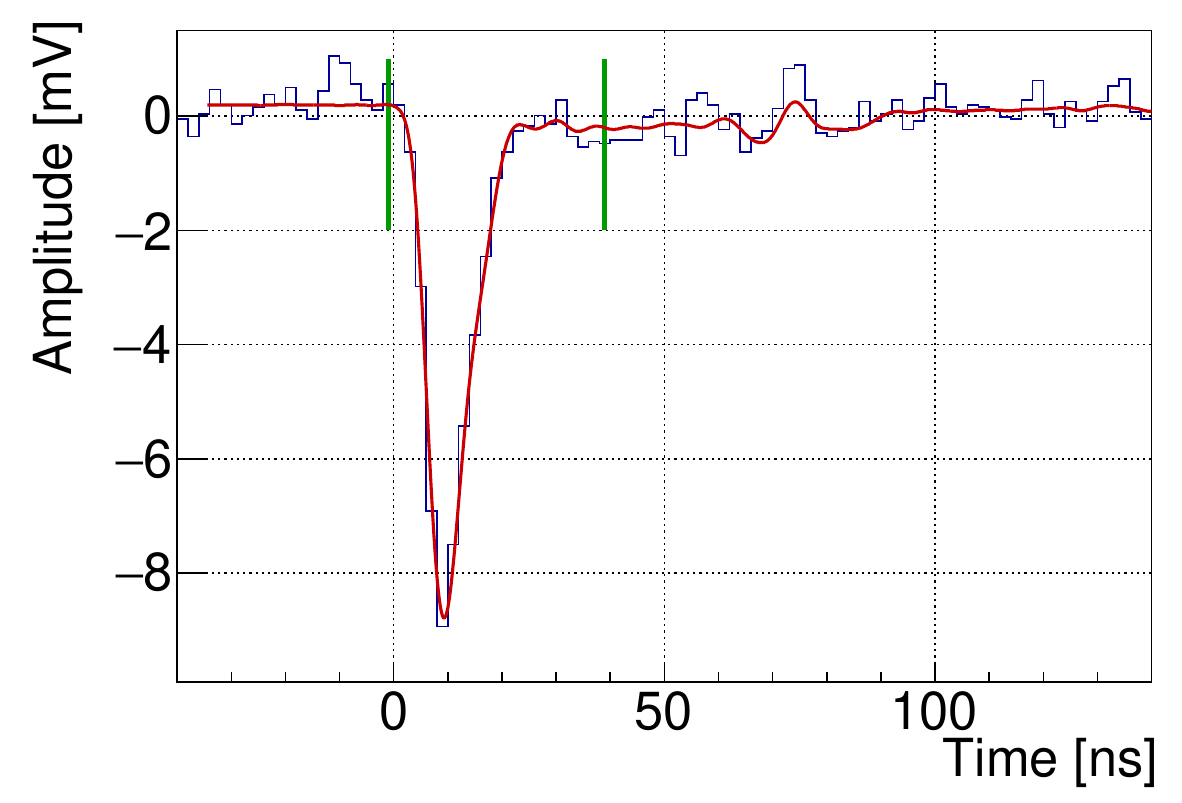}
 \hfill
 \includegraphics[width=0.49\linewidth]{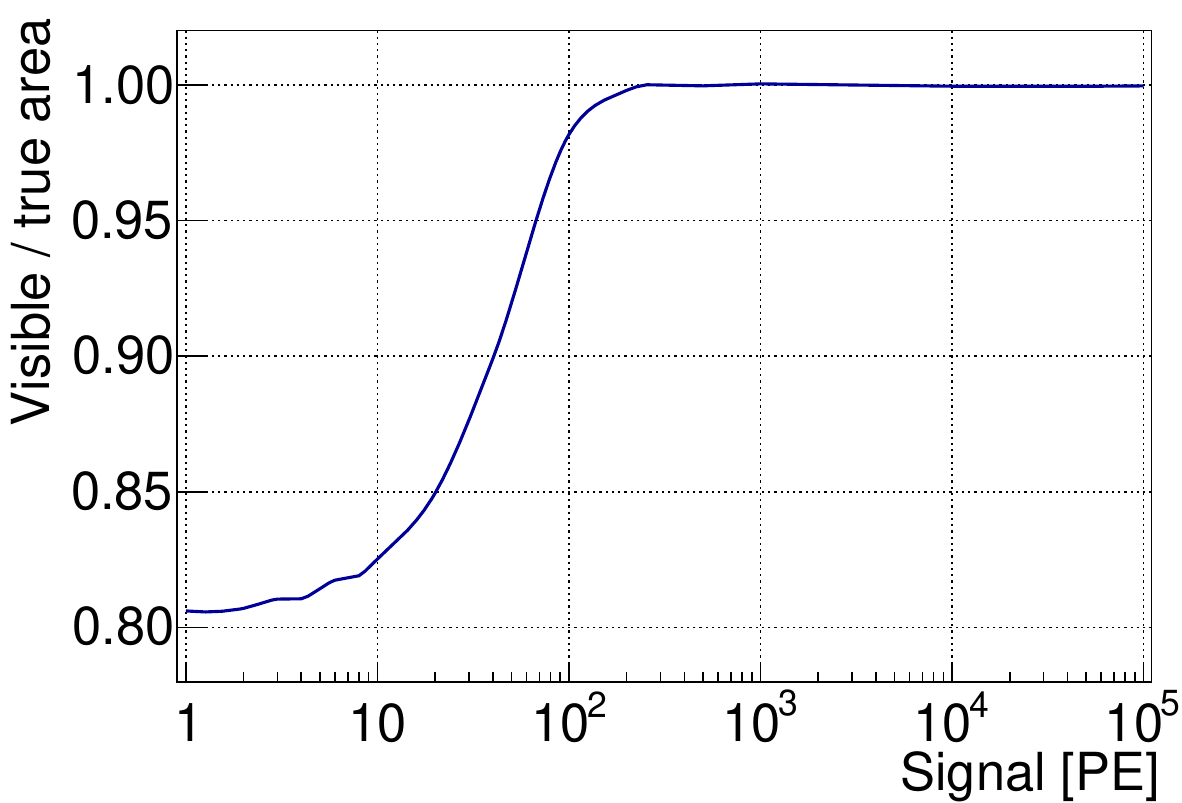}
	\caption{\textbf{Left}: Example of the waveform with SPE signal on it. Green lines indicate borders of the identified pulse. The red line indicates the measured averaged shape of the SPE pulse. \textbf{Right}: The efficiency of the pulse area evaluation depending on the number of PE generated per event.}
  \label{img:spe_shape_eff}  
\end{figure}

The parameters of SPE pulses are evaluated with the help of the LED calibration.
These pulses are defined as pulses with a signal amplitude of more than 2 mV and a duration (full width at half amplitude) of more than 4~ns. 
This unified amplitude threshold is used since PMT voltages are tuned to keep SPE amplitudes approximately the same for all channels. 
As one can see in the example waveform shown in figure \ref{img:spe_shape_eff} (left), some part of the SPE pulse is outside of the determined pulse time window width, which is mentioned in the previous paragraph. 
The red line on the plot shows the average shape of SPE signal for the individual PMT and it lies below the baseline level in the region after the right defined border of the pulse.
It is a complicated problem to account for this part since the amplitude in this region is very close to the noise level.
The ordinary extension of the integration region leads to the degradation of resolution and pulse separation.
 This feature of the signal introduces an additional source of systematic bias in SPE area calculation. 
In the case of large signals, when individual SPE signals are combined into one big pulse, these trailing pieces are included in the total sum.
To estimate the area losses, the toy Monte-Carlo simulation is performed. We measure the average SPE signal shape using a high sampling rate oscilloscope. 
Then, a known number of SPE pulses, distributed uniformly in 2~$\mu$s (a characteristic SE signal), is added to the real waveforms without detected pulses. 
The number of the SPE pulses is varied from 1 to $10^6$. 
The constructed waveforms are then processed by REDOffline as usual.
As shown in figure~\ref{img:spe_shape_eff}, right, there is a significant decrease in area calculation efficiency for SE-like signals represented by sparse SPE pulses. 
The asymptotic value of this efficiency for SPE is 81$\pm$1\%, which coincides with the result obtained using direct integration of SPE signal shape using average pulse time window width for SPE pulses. 

\begin{figure}[htbp]
  \begin{minipage}[ht]{0.49\linewidth}    \center{\includegraphics[width=1.0\linewidth]{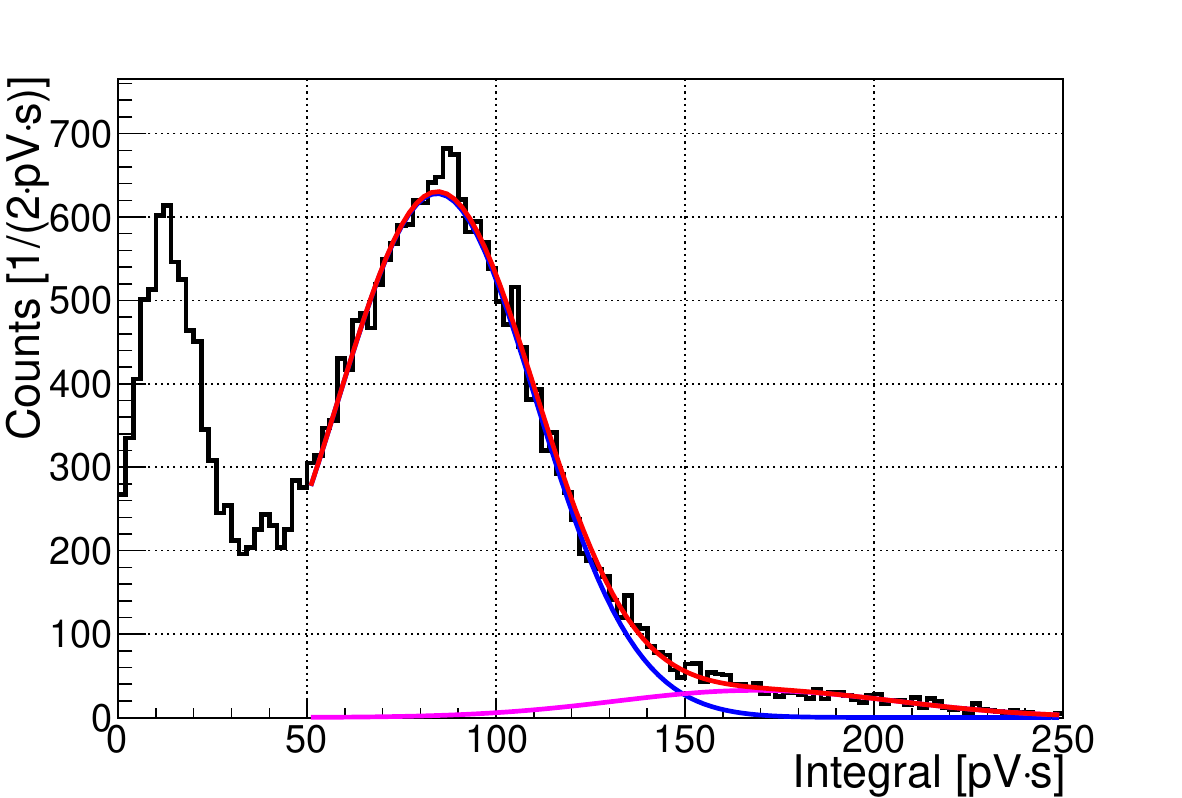} \\}
  \end{minipage}
  \hfill
  \begin{minipage}[ht]{0.49\linewidth}  \center{\includegraphics[width=1.0\linewidth]{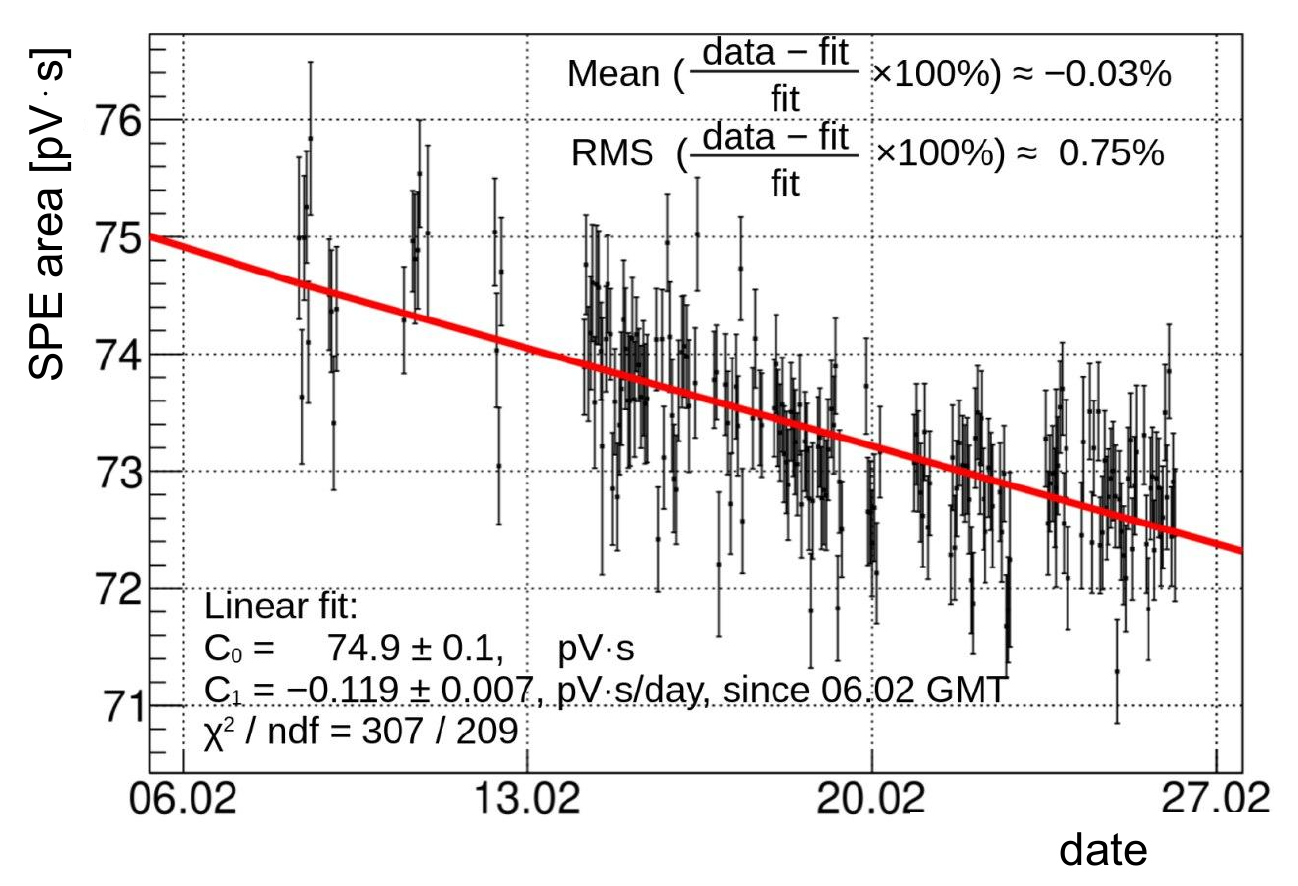} \\}
  \end{minipage}
    \caption{\textbf{Left}: Example of the SPE spectrum for PMT T09 (top array) and the result of the fit (red line) with the contributions of SPE (blue line) and DPE (magenta line). \textbf{Right}: Dependence of the mean SPE integral on time for PMT T09 (top array) and the result of the fitting (red line).}
\label{img:spe2022}
\end{figure}

An example of the pulse area distribution is shown in figure~\ref{img:spe2022}, left.
As one can see, the SPE peak is well-separated from the pedestal, thereby permitting fitting by two Gaussians (for single and double SPE pulses). 
The second peak can be caused by double photo-electron (DPE) emission~\cite{Faham_2015}, or simple coincidences.
The mean SPE value for each channel, derived from the fit, is used to quantify the areas of all other signals in PE units using the efficiency coefficient described below. 
During the scientific data collection period, slow changes of the mean SPE area are observed. 
It is accounted for by fitting the time dependence of the mean SPE area with a linear function. Typical example for one of the channels is shown in figure~\ref{img:spe2022}, right.
Despite many channels having moderate values of $\chi^2/ndf$, the mean deviation of data from the fit does not exceed 0.1\% and the root mean square of this deviation is less than 0.9\%. 
This variation has a negligible effect on the calibration results.
In further analysis, the value of this function for a particular data-taking run is used as the SPE size.

\subsection{Electron lifetime}
\label{subsec_lifetime}
We evaluate the lifetime of a free electron in liquid xenon using signals from cosmic muons. This approach is described in detail in ref.~\cite{The_RED100_Experiment, lifetime} and was shown to yield a result similar to a conventional approach relying on signals from gamma rays~\cite{7027252}. Here we provide only a brief overview of the former method.

\begin{figure}
    \centering
    \includegraphics[width=0.6\linewidth]{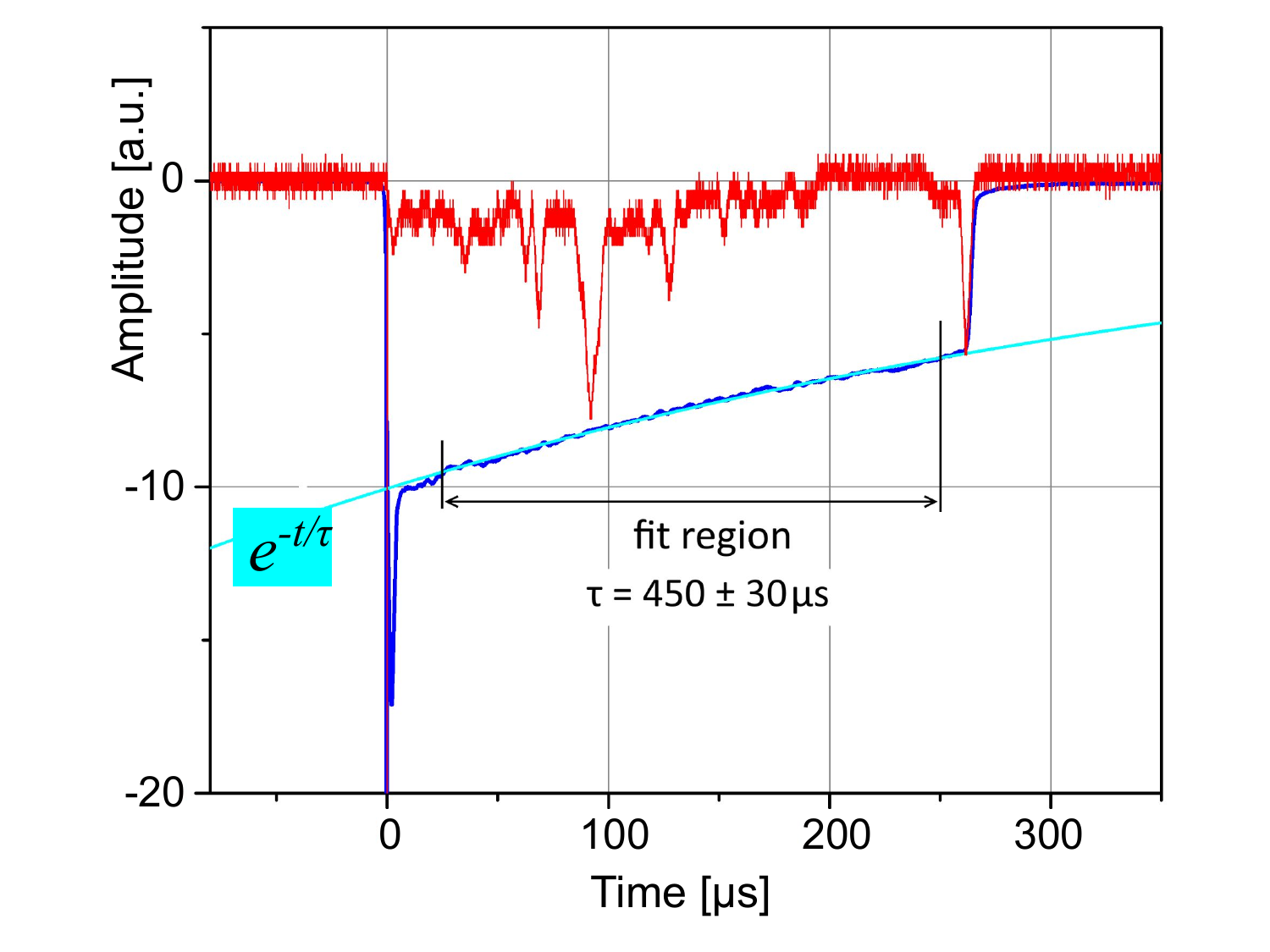}
    \caption{The individual muon signal (red) and the average signal (blue) with the exponential fit (cyan).}
    \label{fig:muon}
\end{figure}

Atmospheric muons are energetic and pass through the detector, leaving a continuous ionized track along the path. 
When many individual muon signals are summed up, the resulting signal represents the average $dE/dx$, which is constant throughout the detector depth. 
The examples of an individual muon signal and the average signal are shown in figure~\ref{fig:muon}.
The electron drift speed does not depend on the source of ionization or deposited energy. 
The S2 size versus depth follows exponential decay because of finite electron lifetime. 
Hence, it is possible to obtain lifetime using the average muon waveform by fitting it with the exponential function as it is demonstrated in figure~\ref{fig:muon}.
The plot illustrating the evolution of the lifetime throughout the RED-100 run at KNPP, measured using cosmic muons, can be found in~\cite{The_RED100_Experiment}.
The average muon waveform also makes possible the calculation of the drift speed by dividing the drift distance by the maximum drift time, which corresponds to the length of the average signal. 

\subsection{Light response functions (LRFs)}
\label{subsec_LRFs}
\begin{figure}[htbp]
    \centering
    \includegraphics[width=0.4\linewidth]{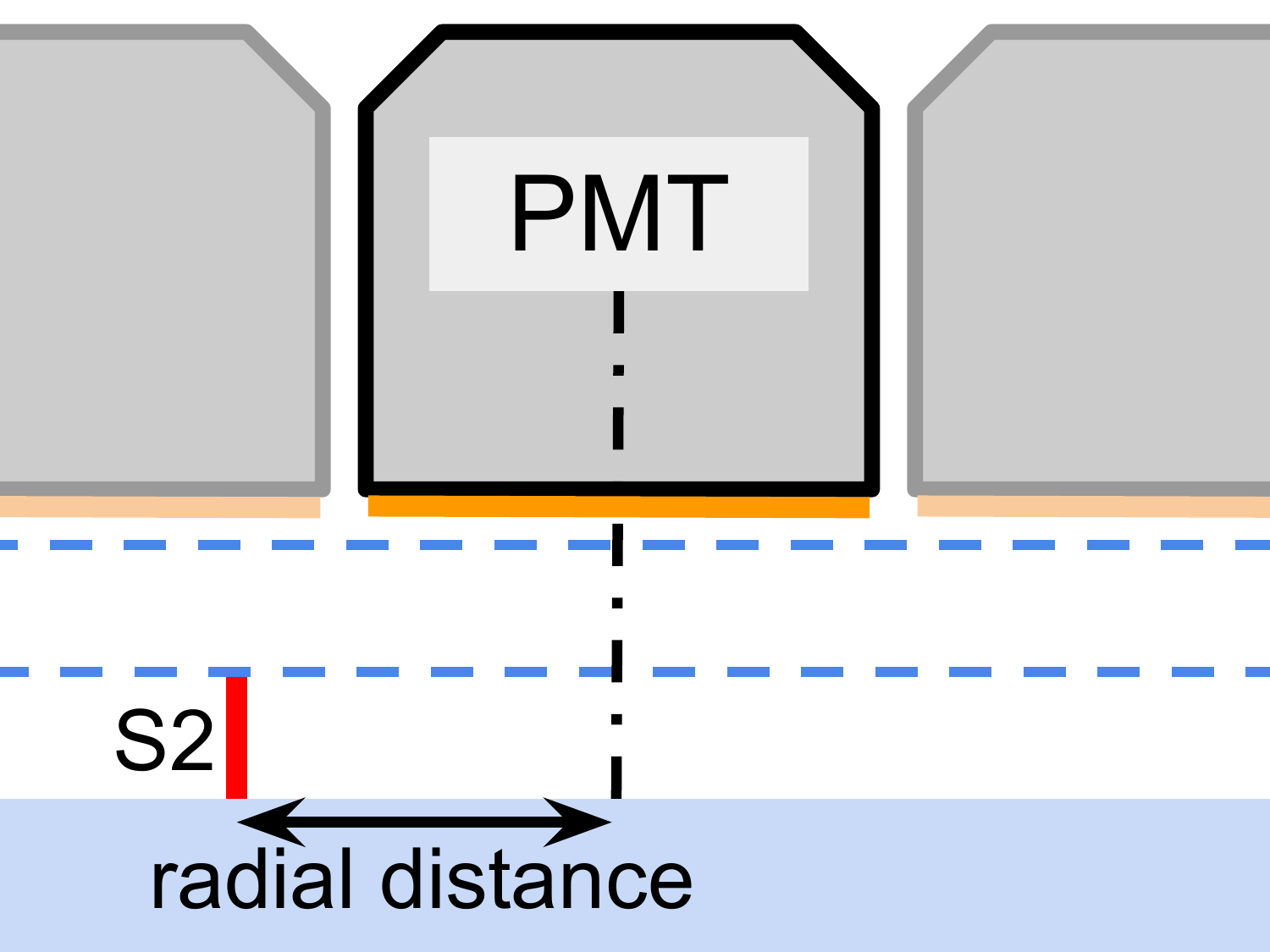}
    \caption{The scheme of the radial distance definition.}
    \label{fig:lrf_scheme}
\end{figure}

There are various methods to account for spatial variations of the light collection efficiency of a two-phase detector. The reconstruction method we choose is based on the use of light response functions (LRFs)~\cite{Solovov2012} for each PMT. LRF is a dependence of the PMT signal on the relative light source position to the center of the PMT's photocathode. 
In simple cases, only one dimension (radius) is considered, which corresponds
to the projection of the distance between the PMT and the light source on the XY plane (see figure~\ref{fig:lrf_scheme}). 
More complex cases utilize two (XY or RZ) or three (XYZ) dimensions. In our analysis, we employ 1-dimensional LRFs. 
All methods involving LRFs operate under the assumption that the LRF shape is independent of the light amount in the linear operation mode of the PMT.
There are several ways to determine the LRF shape, such as Monte-Carlo simulation, analytical calculations, or analysis of experimental data. In this work, the latter one is used based on the calibration with gamma sources.

\subsubsection{Event selection}
\label{subsubsec:event_select}

Analysis of the gamma-calibration data requires the selection of pulse groups (clusters) corresponding to S1 and S2 signals on the waveforms. 
The signals from the full absorption peak for the gamma sources we use are large enough to obtain continuous S1 and S2 pulses above the threshold distributed over many channels. 
In this case, a so-called effective summed waveform can be used for data processing. 
The following algorithm was used to identify S1 and S2 clusters:
\begin{enumerate}
    \item
    The initial pulses are converted to rectangular ones with the constant amplitude and the duration equal to the pulse time region.
    The amplitude of each rectangular pulse is calculated by dividing the primary pulse area by its duration. 
    \item Rectangular pulses obtained in all the channels corresponding to the top PMT array are combined into the effective summed waveform.
    \item 
    The pulses of the effective waveform corresponding to S1 or S2 are identified. Their amplitude and width (which is defined as a full pulse time region) are used to determine pulse type according to the following criteria: a duration between 140 and 340~ns and an amplitude between 0.2 and 0.9~V for S1,
    a duration of more than 1700~ns and an unrestricted amplitude for S2. The clusters with a duration between 340 and 1700~ns are nontypical clusters and are not selected for further analysis.

\end{enumerate}
Afterward, the so-called full events are constructed using the identified S1 and S2 clusters. The events with only one S1 and only one S2 are selected for analysis.

\begin{figure}[htbp]
  \begin{minipage}[ht]{0.49\linewidth}    \center{\includegraphics[width=1.0\linewidth]{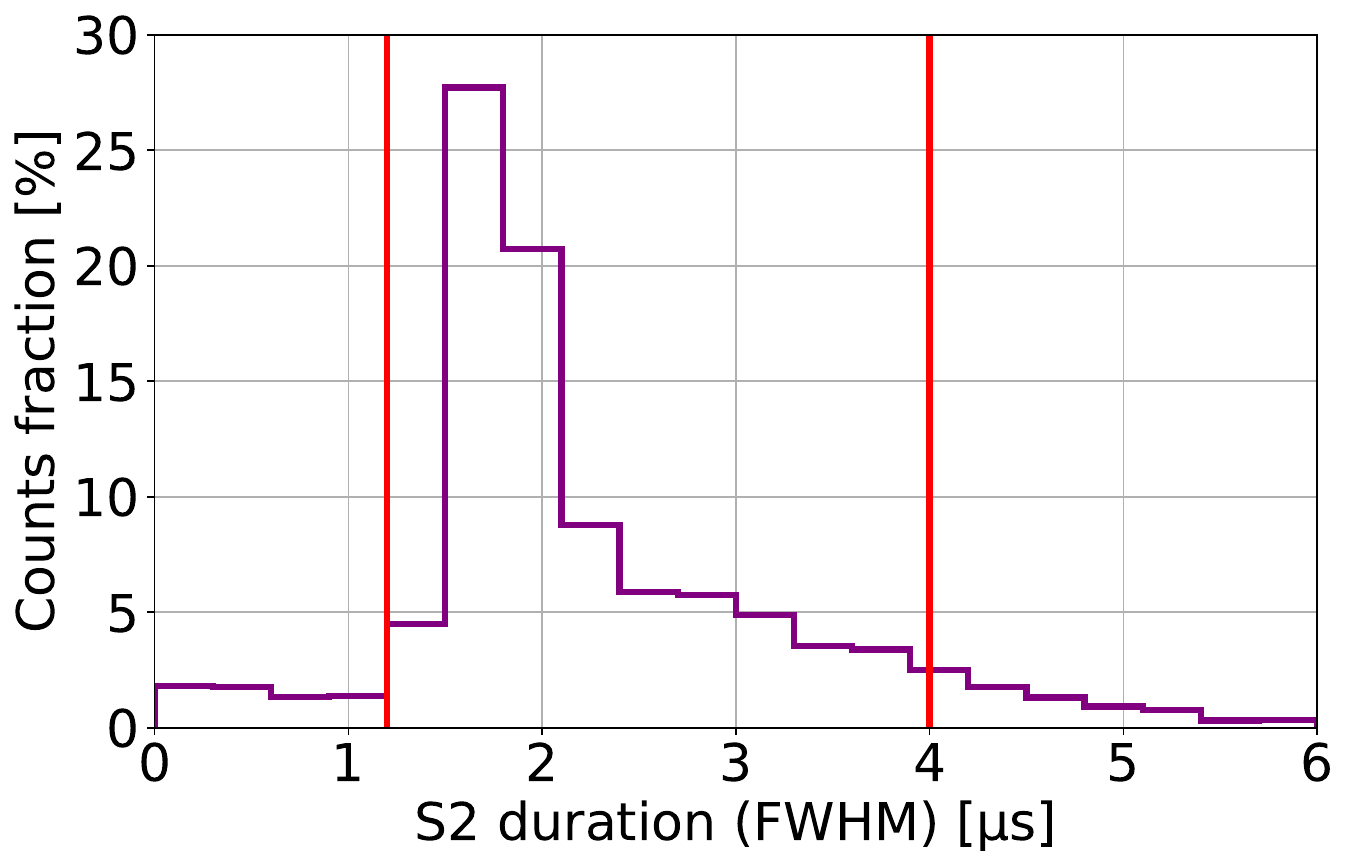} \\}
  \end{minipage}
  \hfill
  \begin{minipage}[ht]{0.49\linewidth}  \center{\includegraphics[width=1.0\linewidth]{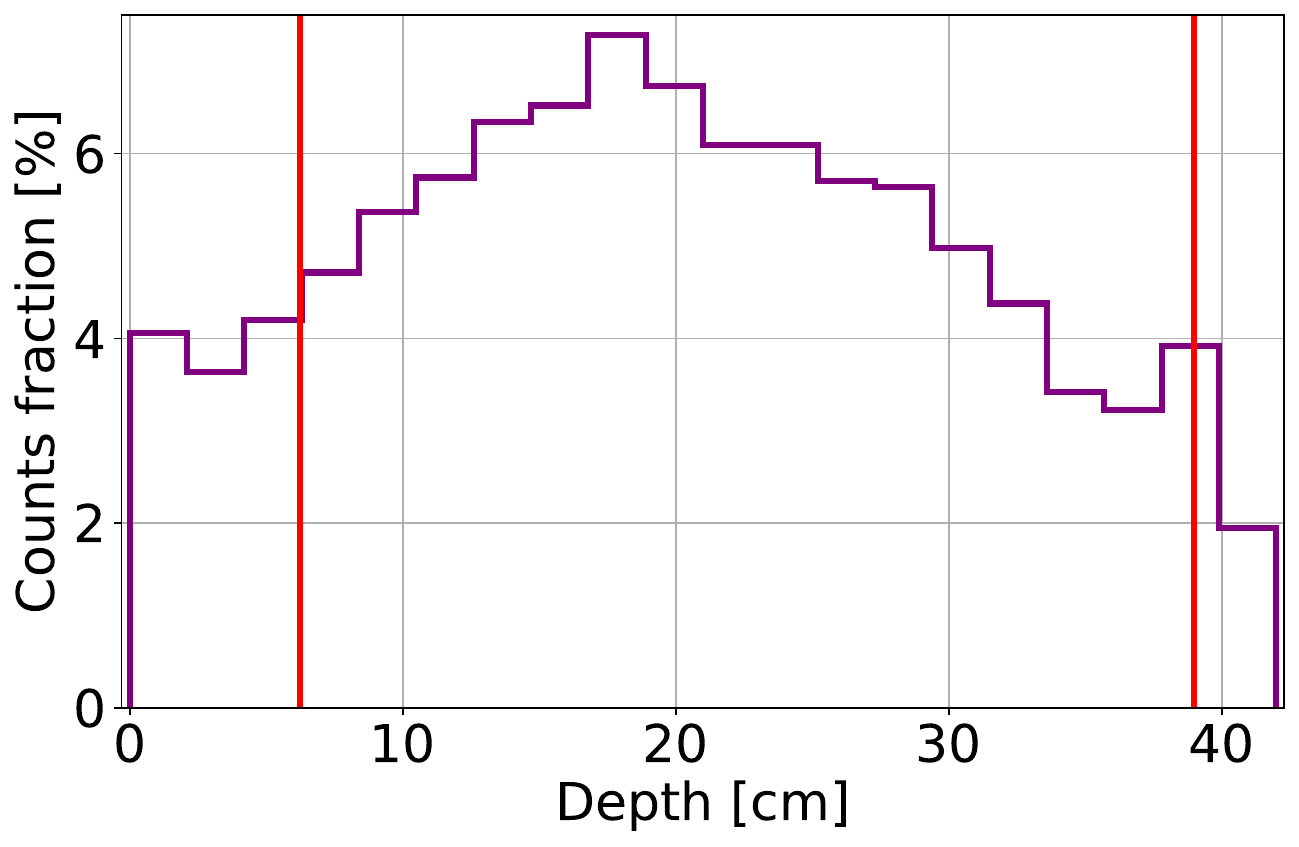} \\}
  \end{minipage}
	\caption{\textbf{Left}: The example of S2 duration distribution for the run with gamma source $^{60}$Co  with the corresponding cut (1.2 -- 4 $\mu$s) (red). \textbf{Right}: The example of depth distribution for the run with gamma source $^{60}$Co with the corresponding cut (40 -- 250 $\mu$s of the drift time which corresponds to approximately 6 -- 39 cm) (red).}
	\label{img:depthdur}
\end{figure}

A two-phase detector technique enables 3-D reconstruction.
The event depth (corresponding to the Z coordinate) can be reconstructed using the time interval between S1 and S2, whereas the XY (horizontal plane coordinates) and energy reconstruction requires a more complex procedure.
The correction function is measured using $^{137}$Cs data. 
As mentioned earlier, the S2 signal is depth-dependent due to ionization electron losses. 
The S2 area in each channel is corrected using the electron lifetime obtained from the muon data by the exponential function of depth:
\begin{equation}
A_{{i }}=A_{i}^{\text{raw}} \cdot e^{{t_{\text {drift}}}/{\tau}},
\end{equation}
where $A_{i}^{\text{raw}} $ represents the S2 area in $i$-th channel before the correction, $A_{{i }}$---after the correction, $\tau$---the measured electron lifetime, $t_{\text {drift}}$---the time difference between S1 and S2 onsets. 
Then, the depth and duration cuts are applied. 
The duration of the S2 pulse is measured as full width at half maximum (FWHM) of the original pulse to minimize the influence of the diffusion.
Examples of event distributions on the depth and the duration, along with the corresponding cuts, are shown in figure~\ref{img:depthdur}.

\begin{figure}[htbp]
  \begin{minipage}[ht]{0.49\linewidth}    \center{\includegraphics[width=1.0\linewidth]{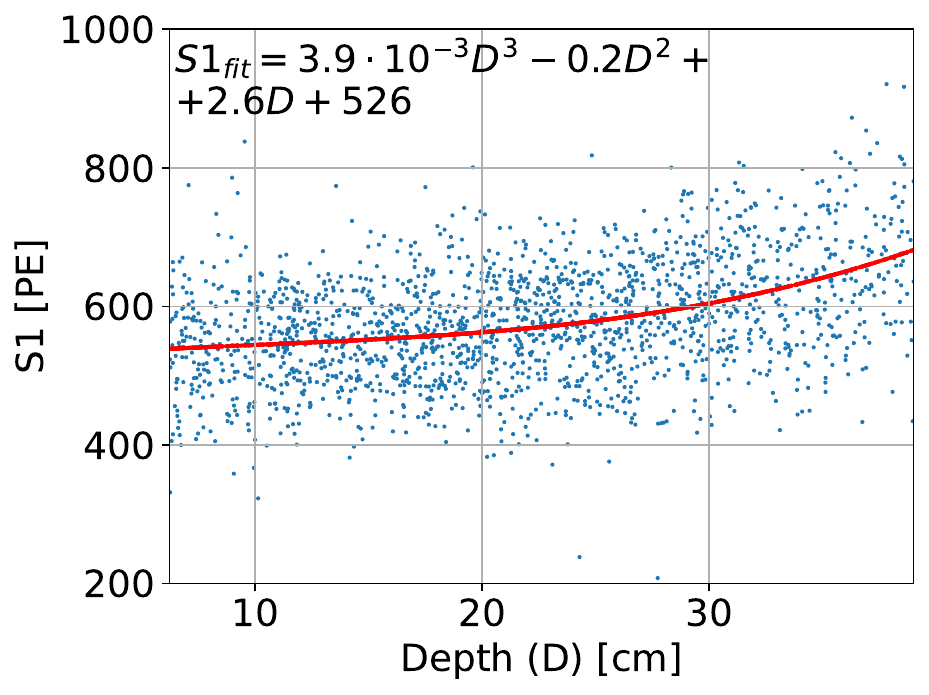} \\}
  \end{minipage}
  \hfill
  \begin{minipage}[ht]{0.49\linewidth}  \center{\includegraphics[width=1.0\linewidth]{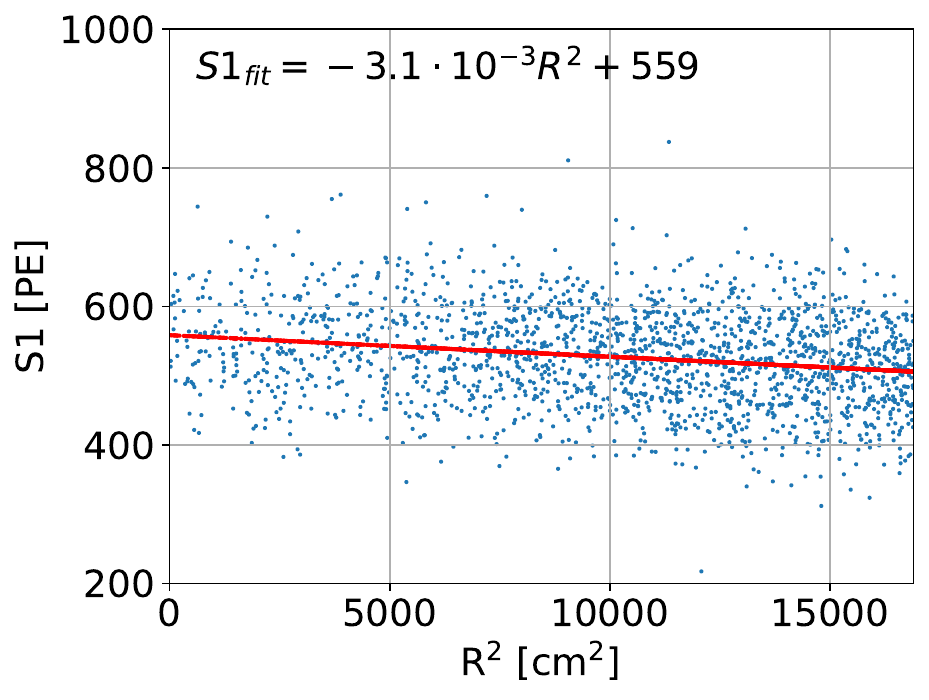} \\}
  \end{minipage}
	\caption{\textbf{Left}: The dependence of the S1 area on the event depth ($^{137}$Cs data) and the result of the fit. \textbf{Right}: The dependence of the S1 area on the radius squared ($^{137}$Cs data) and the result of the fit.}
	\label{img:S1_depend}
\end{figure}

The detector response depends on the event XY position due to the top PMT array light collection efficiency. 
The correction and reconstruction procedure is performed only for S2 signals. 
Detailed correction of the S1 requires 3-D LRFs, for which we don't have a sufficient amount of calibration data.
However, we apply simple two-step correction to the S1 signal area.
In the first step, the dependence on the depth is corrected using the cubic polynomial.
The S1 signal depends on the interaction depth because the events closer to the detector's bottom give a greater light response.
This is because the bottom PMT array is more efficient in detecting the S1 signal than the top one due to the total reflection from the surface of a liquid. 
In the second step, the dependence on the radius squared was taken into account using a linear function. The radius is reconstructed with a procedure described below.
The corresponding coefficients of the correction functions were evaluated by fitting the data selected from the $^{137}$Cs peak as illustrated in figure~\ref{img:S1_depend}.
We use $^{137}$Cs peak data because it has a mono-energetic line.

\subsubsection{LRF evaluation}
\label{subsubsec:LRFeval}
For the LRF reconstruction procedure, we use the Mercury~\cite{Solovov2012} iterative algorithm on $^{60}$Co experimental data. 
The environment for the LRF evaluation procedure is implemented in ANTS2~\cite{Morozov_2016}, a software package designed for the position, energy reconstruction, and optical modeling of detectors. 
It also includes the implementation of the LRF fitting and various reconstruction methods, one of which was used in our analysis. The specific implementation of it is described below.

\begin{figure}[htbp]
  \begin{minipage}[ht]{0.43\linewidth}    \center{\includegraphics[width=1.0\linewidth]{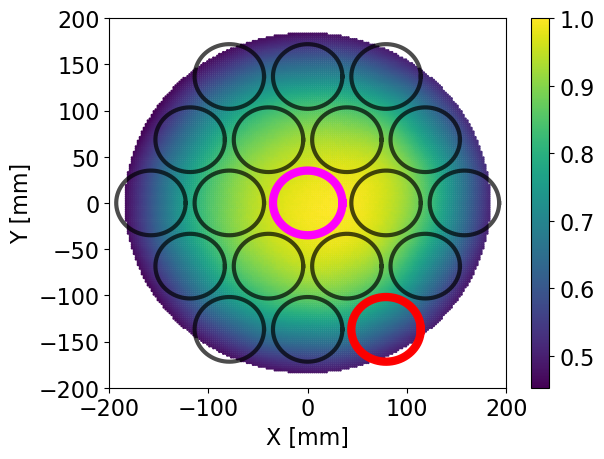} \\ }
  \end{minipage}
  \hfill
  \begin{minipage}[ht]{0.55\linewidth}  \center{\includegraphics[width=1.0\linewidth]{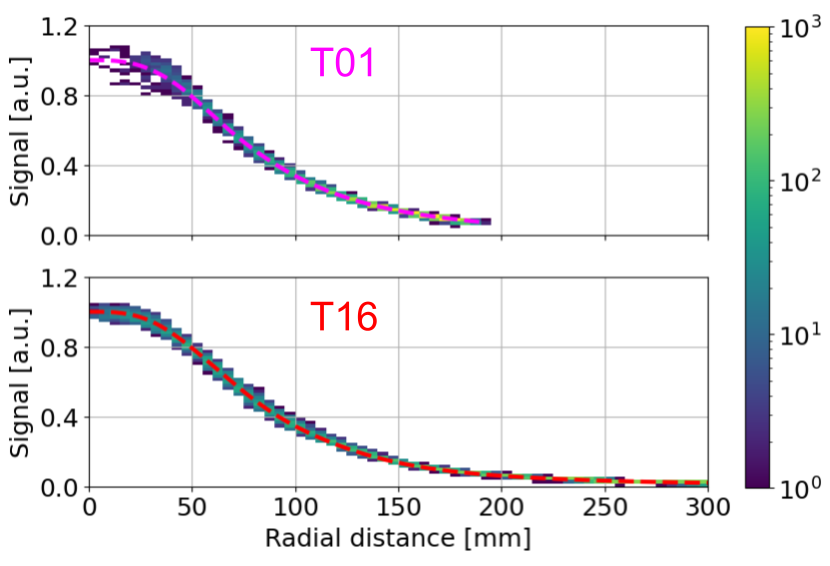} \\ }
  \end{minipage}
	\caption{\textbf{Left}: The light collection efficiency map. \textbf{Right}: The example of the reconstructed LRF (red line) and signal distribution for PMTs T01 and T16 (PMTs are indicated with corresponding colors in the left figure). The signal is  normalized by the energy calculated using the formula~\ref{Energy_formula} and to the maximal values of LRFs.}
	\label{img:xy_and_lrf}
\end{figure}
\begin{figure}[htbp]
    \center{\includegraphics[width=1.\linewidth]{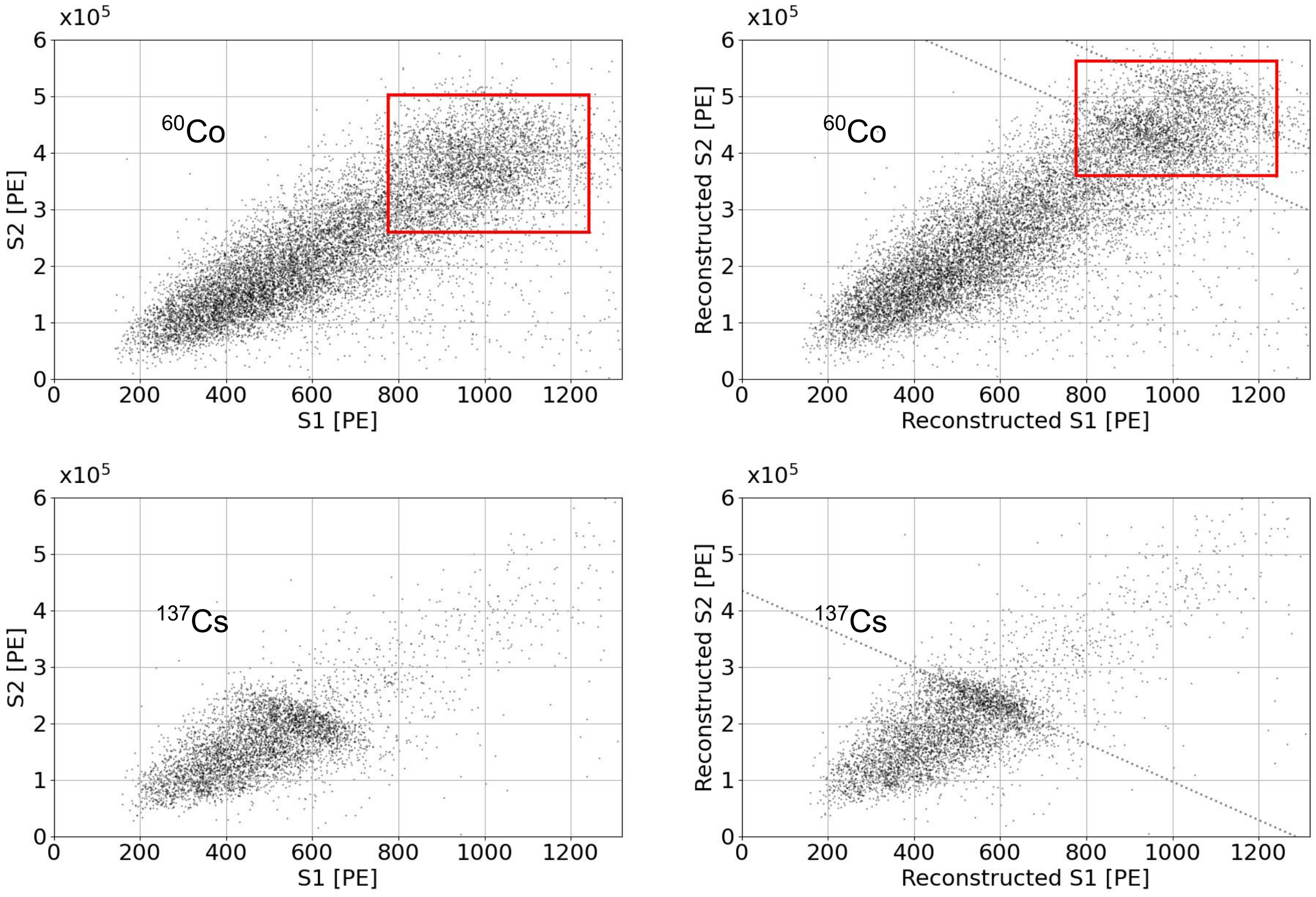} \\}
	\caption{\textbf{Left}: S2 vs S1 scatter plot before reconstruction.
 \textbf{Right}: S2 vs S1 scatter plot after reconstruction. \\The $\chi^2$ and the R cuts are applied to the events on both distributions. The red lines show selection of the events taken into LRF reconstruction. The grey dotted lines indicate the selected $^{60}$Co and $^{137}$Cs peaks.}
  \label{img:2dspectra}  
\end{figure}
On each iteration, the distributions of the signal on distance from the PMT's axis are fitted by B-splines to obtain LRFs.

If the Mercury algorithm converges, the result does not depend on the start positions.
In our case, the events from the region of two close photopeaks from 1173 and 1333~keV are preliminary chosen (as shown in figure~\ref{img:2dspectra}, top-left) on the first iteration, and then they are chosen with greater accuracy for the subsequent iterations of the process (figure~\ref{img:2dspectra} top-right).
The events from peaks are selected because they have well-defined energy.
The coordinates for the first iteration are calculated using the centroid method and then multiplied by an empirical coefficient of 1.8.
The motivation of this coefficient is scaling the edges of the XY distribution reconstructed by the centroid algorithm to the detector size.
The centroid, also known as the center-of-gravity method, is the simplest and fastest method for position and energy reconstruction which does not require LRFs. 
In this method, the event coordinates are determined as follows:
\begin{equation}
X_{\text {event }}=\frac{\sum_i A_i X_i}{\sum_i A_i}, \quad Y_{\text {event }}=\frac{\sum_i A_i Y_i}{\sum_i A_i},
\end{equation}
where ($X_i$, $Y_i$) represent the coordinates of the $i$-th PMT's axis, and $A_i$ is the measured area of the S2 signal from this PMT.
For subsequent iterations, the maximum likelihood estimation method with contracting grids~\cite{grids} is applied for XY reconstruction of all events. 
The likelihood function was computed as follows:

\begin{equation}
L_{\text {event }}=-\sum\limits_{i=1}^{19} \Big( A_i \ln(LRF_i(x,y) A_{\text {event}}^{\text {corr}}) - LRF_i(x,y) A_{\text {event}}^{\text {corr}} \Big),
\end{equation}
where $A_i$ is the measured S2 signal area from the $i$-th PMT of the top array, $LRF_i(x,y)$ is the LRFs value from $i$-th PMT corresponding to the reconstructed coordinates $x,y$, and $A_{\text {event}}^{\text {corr}}$ represents the corrected S2 area, which is calculated as:

\begin{equation}
\label{Energy_formula}
A_{\text {event}}^{\text {corr}}={\sum\limits_{i=1}^{19} A_i}\Big{/}{\sum\limits_{i=1}^{19} LRF_i(x,y)},
\end{equation}

 The S2 versus S1 scatter plots before and after the reconstruction with corresponding event selection are shown in figure~\ref{img:2dspectra}.
Reconstructed S2 energy in PEs is calculated as the energy from~\ref{Energy_formula} multiplied by the coefficient between reconstructed and measured S2. 
This coefficient is calculated using the events from the central area of the detector. 
The distributions presented in figure~\ref{img:2dspectra} are obtained after applying the sum $\chi^2$ and the radius cuts, which are described below. 
An example of the measured LRF is shown in figure \ref{img:xy_and_lrf} (right). 

\subsubsection{Event reconstruction}
\label{subsubsec:rec_res}

The gamma-calibration data is used to build LRFs because it is the most confident collected calibration data. 
In further analysis, the reconstruction based on these LRFs was applied to all other data types.

\begin{figure}[htbp]
  \begin{minipage}[ht]{0.49\linewidth}    \center{\includegraphics[width=1.0\linewidth]{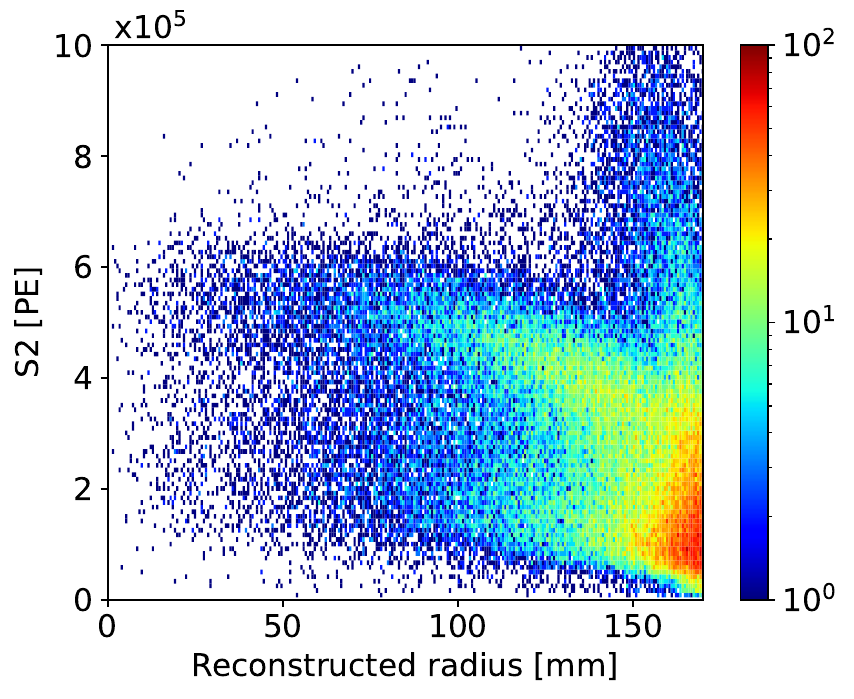} \\}
  \end{minipage}
  \hfill
  \begin{minipage}[ht]{0.49\linewidth}  \center{\includegraphics[width=1.0\linewidth]{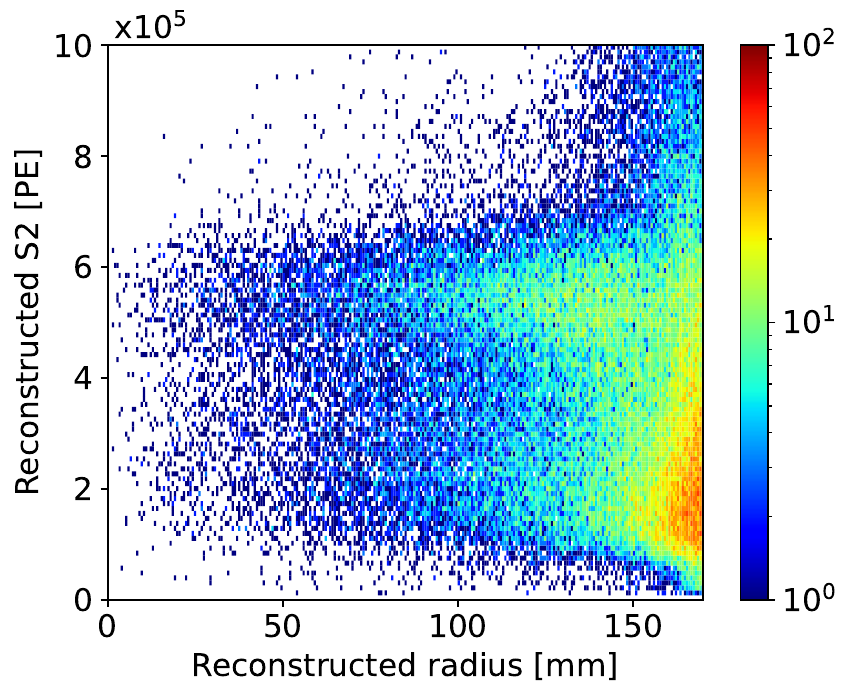} \\}
  \end{minipage}
	\caption{\textbf{Left}: Dependence of the top array sum signal on the reconstructed radius for $^{60}$Co calibration events.
 \textbf{Right}: Dependence of the reconstructed S2 energy on the reconstructed radius for $^{60}$Co calibration events.}
	\label{img:energy_vs_r}
\end{figure}

\begin{figure}[htbp]
  \begin{minipage}[ht]{0.45\linewidth}    \center{\includegraphics[width=1.0\linewidth]{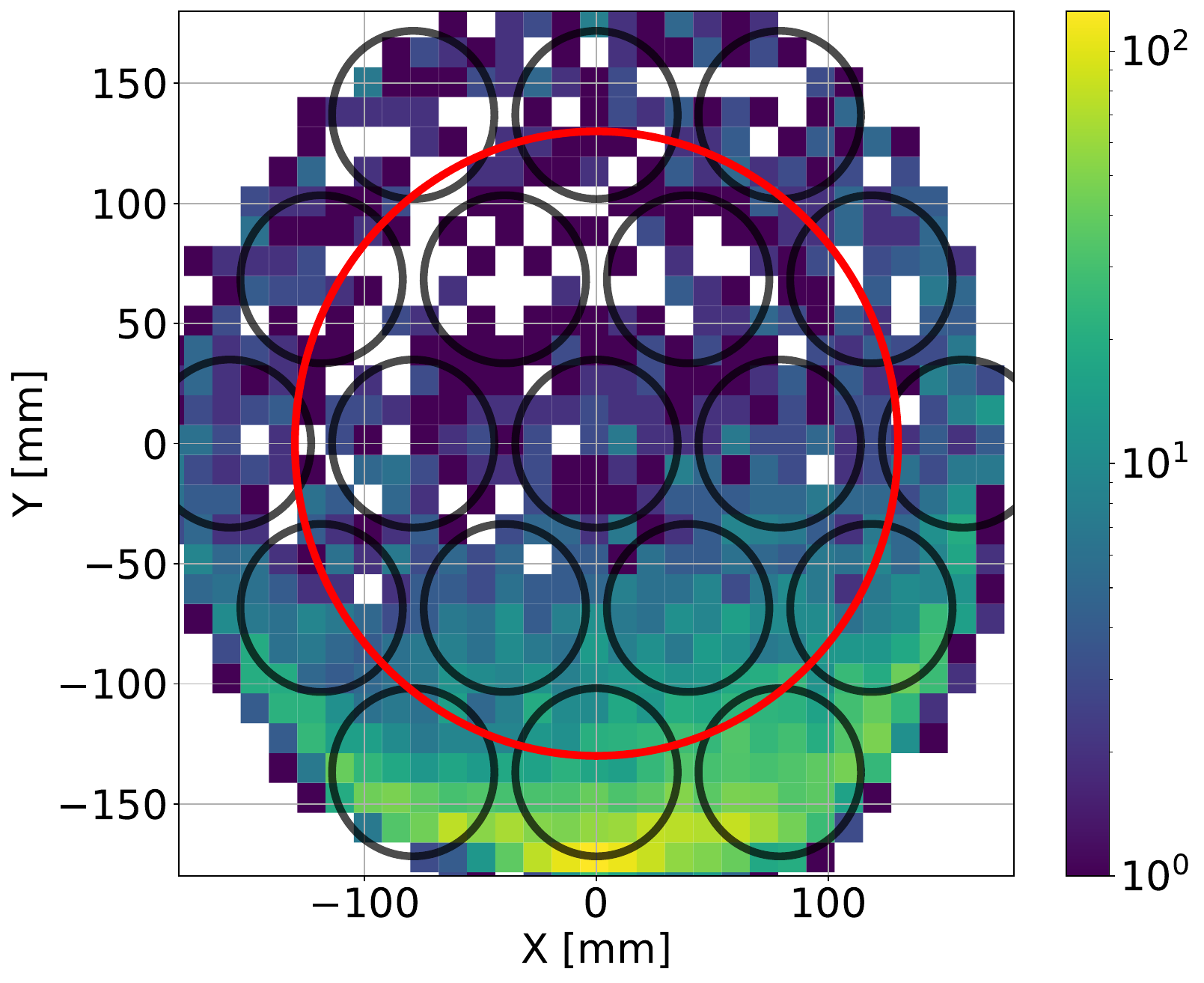} \\}
  \end{minipage}
  \hfill
  \begin{minipage}[ht]{0.53\linewidth}  \center{\includegraphics[width=1.0\linewidth]{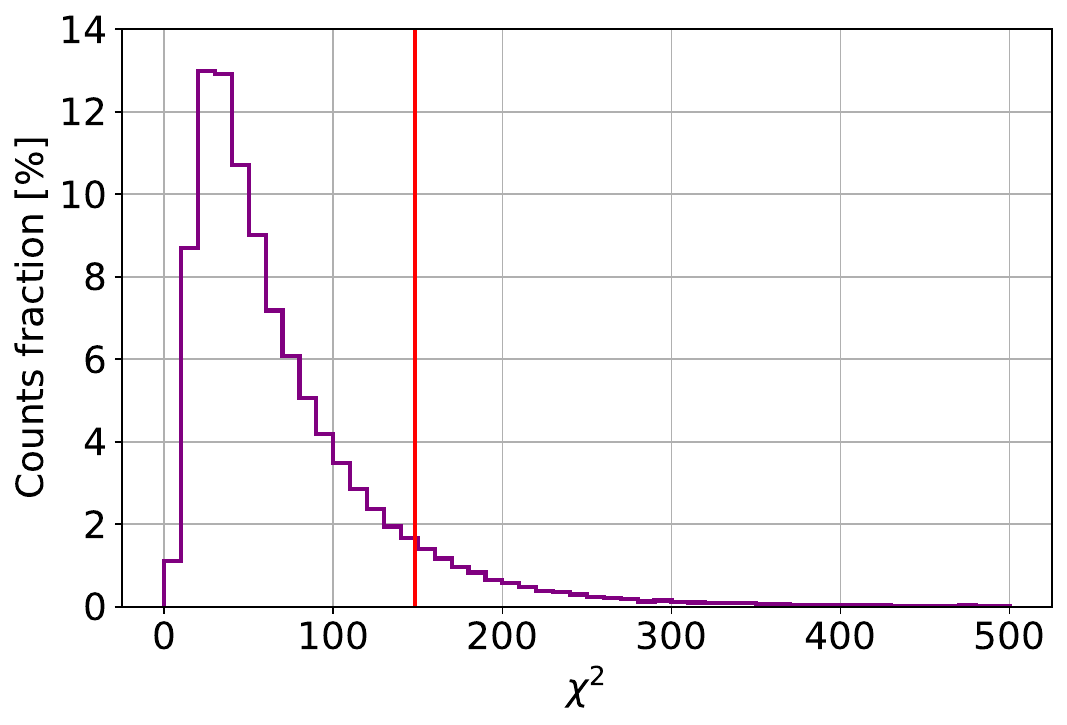} \\}
  \end{minipage}
	\caption{\textbf{Left}: The example of reconstructed XY distribution for calibration source located at one of the points shown in figure~\ref{img:red100geometry}. The red line illustrates the selected cut (130 mm).  \textbf{Right}: $\chi^2$ distribution of all events from $^{60}$Co. The red line indicates the selected $\chi^2$ cut (148.3), which corresponds to the 90-th percentile.}
	\label{img:xy_and_chi2}
\end{figure}

The result of the reconstruction for $^{60}$Co data is shown in figures~\ref{img:2dspectra} (right),~\ref{img:energy_vs_r},~\ref{img:xy_and_chi2} (left).
Additional cuts on the reconstructed radius and sum $\chi^2$ between expected and observed response of PMTs are used to eliminate poorly reconstructed events. 
The $\chi^2$ is calculated using the following formula:
\begin{equation}
\chi^2_{\text {event }}=\sum\limits_{i=1}^{19} \frac{(A_i - LRF_i(x,y) A_{\text {event}}^{\text {corr}})^2}{LRF_i(x,y) A_{\text {event}}^{\text {corr}}}
\end{equation}
An illustration of radius and $\chi^2$ cuts is shown in figure~\ref{img:xy_and_chi2}. 
Two peaks from $^{60}$Co are not resolvable on the spectrum before the position and energy reconstruction (figure~\ref{img:2dspectra}, left).  
The spatial uniformity of S2 signals is corrected based on the S2 light collection correction derived from LRF.
The application of the reconstruction algorithm on the calibration data reduces this dependence and allows us to resolve the two $^{60}$Co peaks (figure \ref{img:2dspectra}, right).

\subsection{SE (single electron) parameters}
\label{subsec:SE}
Spontaneous emission of SE taking place in LXe two-phase detectors~\cite{Akimov:2012zz,Akimov_2016_delayed_electrons, PhysRevD.102.092004, Kopec_2021} can be used for the single electron gain (SEG) evaluation.
Unlike events from gamma sources, which have relatively high energies and have large solid pulses corresponding to S1 and S2, a signal from SE represents the minimal size of electroluminescence and is composed of a few dozen SPE pulses uniformly distributed over the entire length of the electroluminescence.
Consequently, a clustering procedure different from that described in section~\ref{subsubsec:event_select} is required.
The SE signal clusters are identified on the waveforms as groups of SPE pulses using the algorithm detailed below:

\begin{itemize}
    \item 
    The pulses with an area larger than the threshold area are selected. 
    The threshold is calculated as two standard deviations down from the mean value of the first Gaussian on the SPE area spectrum. 
    The use of the threshold connected to SPE size instead of the pedestal helps unify the efficiency of the selection relative to SPE in different channels with varying noise levels.
    \item 
    The clusters are formed as groups of pulses from the top PMT array if the time distance between the start times of two consecutive pulses did not exceed 500 ns.
    The duration of a cluster is calculated as the time distance between the onset of its first pulse and the end of the last one.
    \item The clusters having fewer than five pulses were rejected.
\end{itemize}

\begin{figure}[htb]
  \begin{minipage}[ht]{0.46\linewidth}    \center{\includegraphics[width=1.0\linewidth]{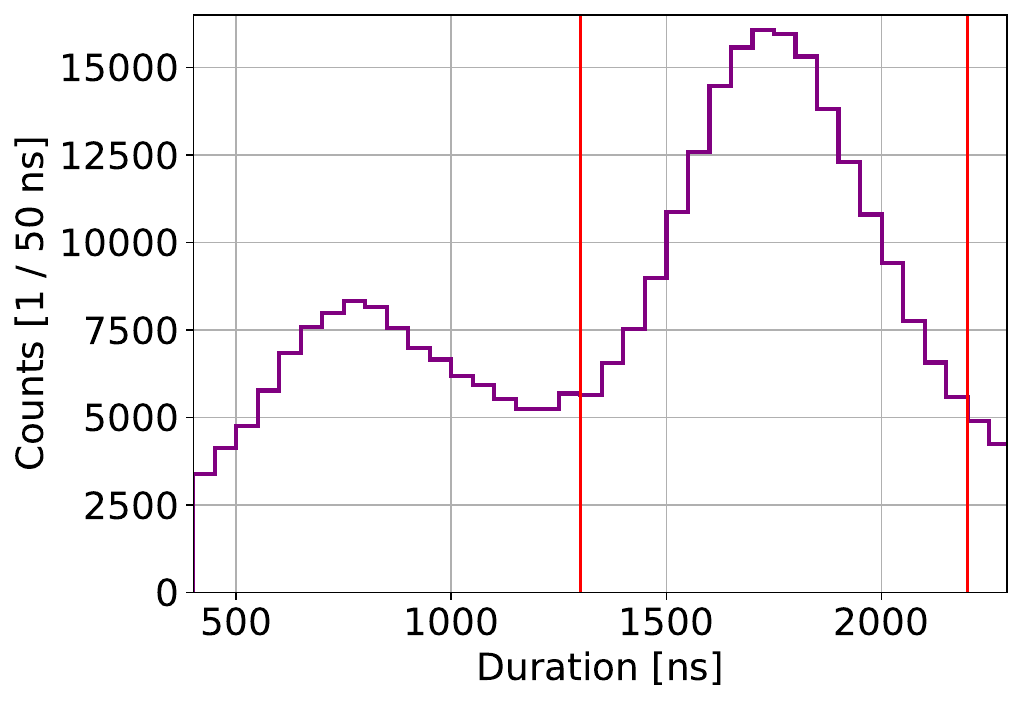} \\}
  \end{minipage}
  \hfill
  \begin{minipage}[ht]{0.52\linewidth}  \center{\includegraphics[width=1.0\linewidth]{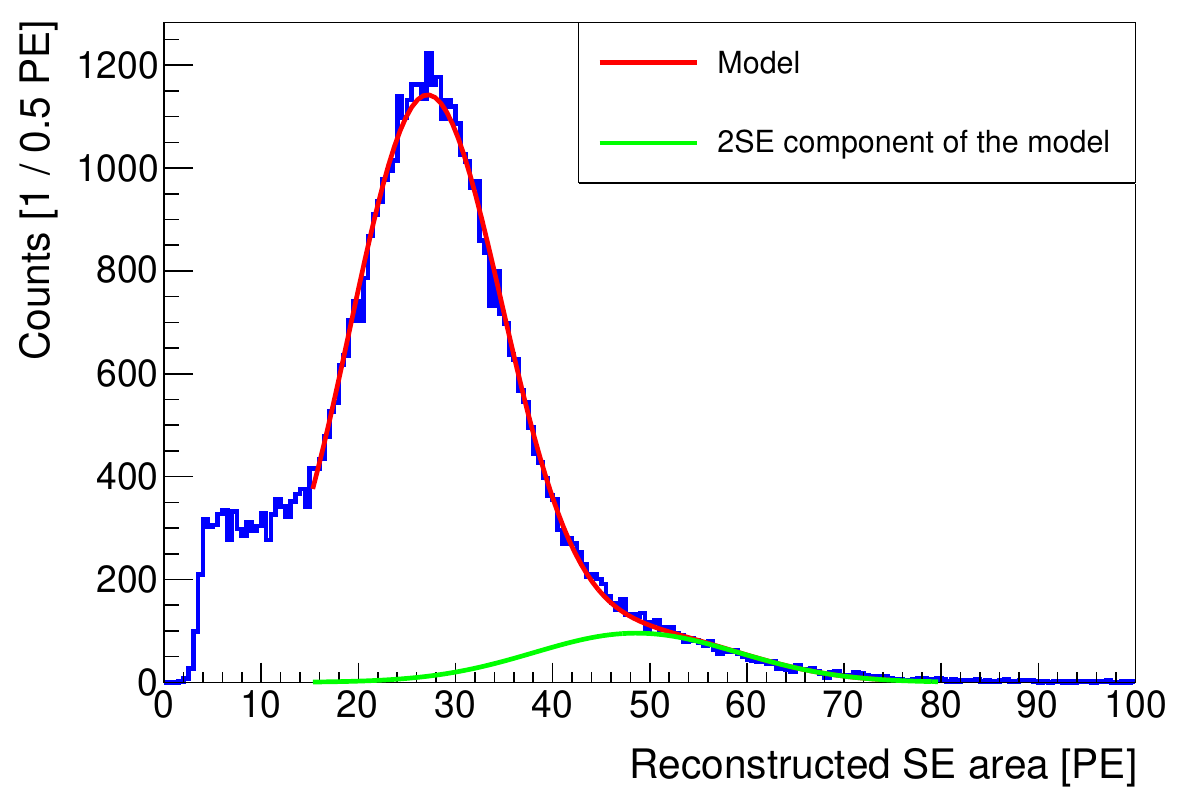} \\}
  \end{minipage}
	\caption{\textbf{Left}: SE duration distribution. \textbf{Right}: The SE area distribution in terms of PEs with the applied cut on the reconstructed radius (<80~mm) with corresponding fit.}
	\label{img:sespectrum}
\end{figure}

After the clusterization, the position and energy reconstruction method described in~\ref{subsubsec:rec_res} is applied to the events. 
The duration distribution of the SE candidate clusters is shown in figure~\ref{img:sespectrum}, left.
We consider, that the peak with a center at $\sim$750~ns corresponds to electron emission at the periphery of the liquid surface due to the peculiarity of the anode electrode design.
The region at the periphery of the detector is excluded from the fiducial volume.

In this work, events in the center (reconstructed radius < 80~mm) of the detector are selected and the reconstructed S2 area spectrum is fitted as shown in figure~\ref{img:sespectrum}. 
SEG measurement in the central region of the detector does not bias the calibration results because the position and energy reconstruction converts the S2 area to the value in the center.

The SE events may coincide with other SEs and with short events from the periphery of the detector and the reconstructed positions of these events are close to the center of the detector. 
Thus, we fit the reconstructed S2 distribution with two Gaussians, where the relative positions of the peaks come from the simple model that takes into account the above-described coincidences with the corresponding estimated probabilities.
The main result of the SE analysis is the single electron gain (SEG), which is equal 27.0$\pm0.1$, as follows from the fit.

\subsection{Electron extraction efficiency (EEE)}
\label{subsec:EEE}
EEE evaluation using the approaches described below requires energy calibration. Calibration of the detector using gamma sources involves obtaining the peaks from monoenergetic lines on the 1-D energy spectrum.
There is an anticorrelation between the S1 and S2 signals in each event~\cite{PhysRevB.76.014115}. 

The reconstruction described in the section~\ref{subsec_LRFs} converts the values of the S1 and S2 areas to the areas in the center of the XY plane near the surface. 
It means, that the influence of the dependence of light collection efficiency on the event position is minimized. Thus, the final energy ($E_{\text{total}}$) can be calculated as a linear combination of the scintillation and the electroluminescence:

\begin{equation}
\label{etotal}
    E_{\text{total}} = W(\frac{S1}{g_{1}}+\frac{S2}{g_{2}}),
\end{equation}
where S1 and S2 are the reconstructed areas, $g_{1}$ and $g_{2}$ are the experimental efficiencies and $W$ is the average energy required to produce an excitation quantum and is equal 13.8$\pm$0.9~eV~\cite{Doke_2002}.
The parameter of the anti-correlation is adjusted for each peak by fitting the peaks on 2-D S1 vs. S2 spectra with 2-D Gaussians.
For $g_{1}$ and $g_{2}$ evaluation we use the average value of anti-correlation parameter.

\begin{figure}[htbp]
  \begin{minipage}[ht]{0.40\linewidth}    \center{\includegraphics[width=1.0\linewidth]{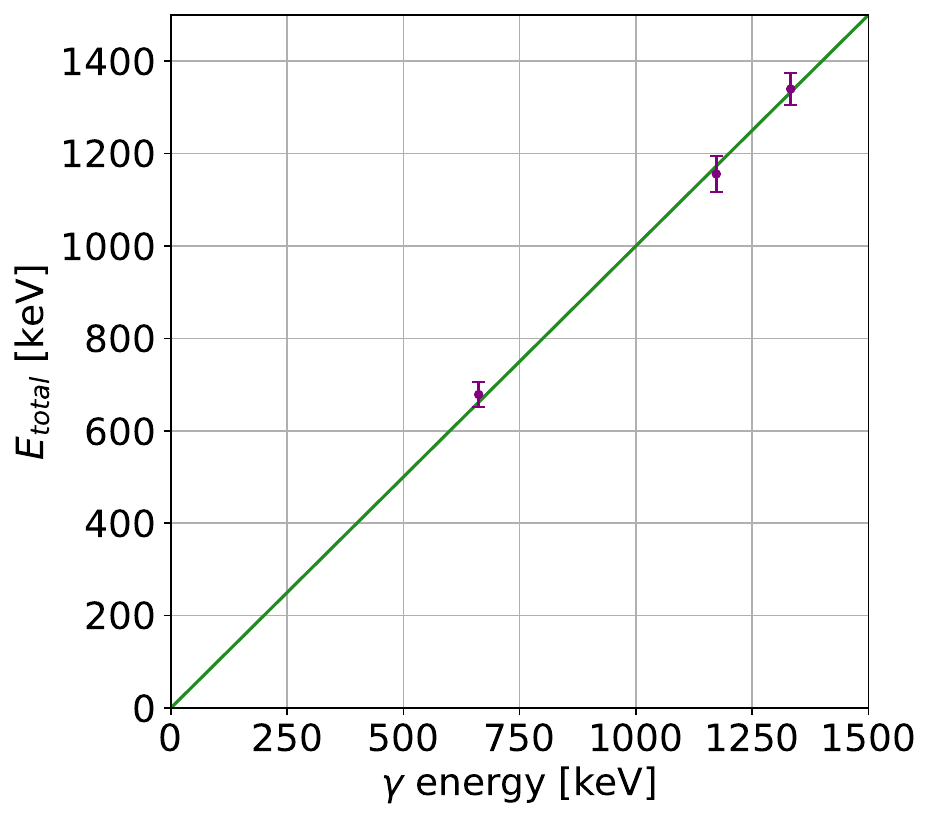} \\}
  \end{minipage}
  \hfill
  \begin{minipage}[ht]{0.58\linewidth}  \center{\includegraphics[width=1.0\linewidth]{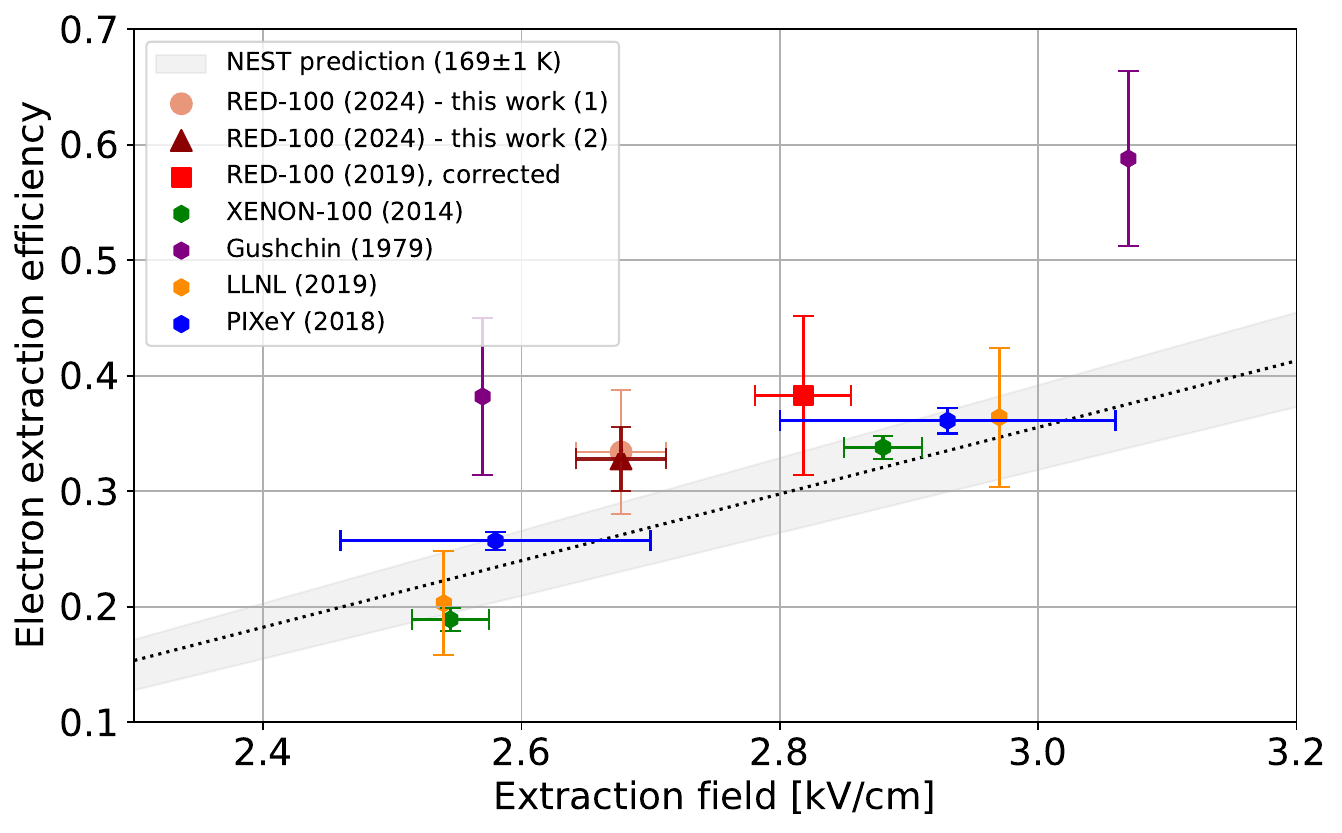} \\}
  \end{minipage}
	\caption{\textbf{Left}: $E_{\text{total}}$ versus the $\gamma$-energy \textbf{Right}: Results of EEE evaluation by RED-100 and other experiments~\cite{Gouschin1978,AprileEEE_2014,Edwards_2018, RED100_2019, PhysRevD.99.103024}.}
	\label{img:EEE_SE_rec}
\end{figure}

The results of the energy calibration are shown in table \ref{tab:calibration_table}.
The obtained $g_{1}$ and $g_{2}$ values are $2.6\pm0.3\cdot10^{-2}$ phd/photon
and $8.8\pm0.8$ phd/electron respectively.
The good linearity of $E_{\text{total}}$ is observed (see figure~\ref{img:EEE_SE_rec}, left). 
The experimental efficiencies ($g_1$ and $g_2$) are comparable to the results obtained by LUX~\cite{LUXg1g2} with a similar detector construction. 
To properly compare $g_1$ values one needs to take into account that only 4 of 19 PMTs in the bottom array of the RED-100 detector are providing data.

\begin{table}[hbt]
    \centering
        \caption{Results of the energy calibration}
\begin{tabular}{|c|c|c|c|c|}
\hline
    $\gamma$ energy, keV2& S1, PE & S2, PE \\
    \hline
    \hspace{0.5em}662  & 574$\pm$18& $(2.40\pm0.04)\cdot10^5$\\
    \hline
    1173  &  942$\pm$27&$(4.26\pm0.08)\cdot10^5$\\
    \hline
    1333 & 1083$\pm$22& $(4.90\pm0.07)\cdot10^5$\\
    \hline
\end{tabular}    
\label{tab:calibration_table}
\end{table}

An electron extraction efficiency (EEE) is defined as the ratio between the number of ionization electrons in the electroluminescence gap after extraction from the liquid ($N_{\text{el}}$) and the original number of ionization electrons in the liquid phase. 
Several approaches to EEE measurements and evaluation can be found in the literature~\cite{Gouschin1978,AprileEEE_2014,Edwards_2018,RED100_2019,PhysRevD.99.103024}. The corresponding results are displayed in figure~\ref{img:EEE_SE_rec}, right. 
We follow two of them in this work.

The first approach relies on the S2 signal only. 
Since the definition of EEE is the ratio between two values proportional to energy, it is possible to switch from the numbers of electrons to the specific values (per keV). 
Thus, in the following calculations, the corresponding specific quantities $Q_{\text{el}} = N_{\text{el}}/E$ and charge yield ($QY$) are used. 
To measure $Q_{\text{el}}$, the S2 values from the table~\ref{tab:calibration_table} are divided by the corresponding $\gamma$ energy and by SEG.

The resulting $Q_{\text{el}}$ in the RED-100 detector for different energies is shown in table \ref{tab:IY_table}. 
To determine the $QY$ value, the NEST (v 2.3.11) \cite{szydagis_m_2023_7577399} package was used. 
The calculated values for the RED-100 drift field of 218~V/cm and density 2.9~g/cm$^3$ are shown in table \ref{tab:IY_table}. Uncertainties of the $Q_{\text{el}}$ value are obtained by combining uncertainties of quantities included in its calculation. 
The resulting EEE value equals 33.4$\pm$5.4\%. 
It is displayed in figure~\ref{img:EEE_SE_rec} (right) as a coral circle (RED-100 (2024) (1)). 
The most significant part of EEE uncertainty in this approach is related to NEST $QY$ prediction accuracy.

The second approach relies on anti-correlation between S1 and S2.
The $g_{2}$ value from eq.~\ref{etotal} equals to EEE $\cdot$ SEG. 
This approach gives the value 32.8$\pm$2.8\% (dark red triangle in figure~\ref{img:EEE_SE_rec}---RED-100 (2024) (2)). The most significant part of the error in this value is related to the W value.

The results obtained using both approaches are in agreement with each other and with other measurements. The second approach gives result with a smaller uncertainty than the first one and is closer to the NEST prediction for the corresponding extraction field.

\begin{table}[hbt]
    \centering
        \caption{Experimental $Q_{\text{el}}$ and $QY$ calculated by NEST. The ratio between $Q_{\text{el}}$ and $QY$ gives the EEE.}
\begin{tabular}{|c|c|c|}
\hline
    $\gamma$ energy, keV & $Q_{\text{el}}$, e$^-$/keV & NEST $QY$, e$^-$/keV\\
    \hline
    \hspace{0.5em}662 & 13.4$\pm$0.4 & 39.7$\pm$6.0\\
    \hline
    1173 & 13.5$\pm$0.3 & 40.6$\pm$6.2\\
    \hline
    1333 & 13.6$\pm$0.3 & 40.8$\pm$6.4\\
    \hline
\end{tabular}    
\label{tab:IY_table}
\end{table}

In our previous analysis in 2019~\cite{RED100_2019}, we used only the first approach and obtained EEE equal to 54$\pm$8\%. The large discrepancy between previous and current measurements was explained by the introduction of the SPE area correction coefficient in this work (see section~\ref{subsec_SPEparams}, figure~\ref{img:spe_shape_eff}). Taking into account this coefficient and the updated NEST charge yield (39.7$\pm$6.0~e$^-$/keV instead of 35.6$\pm$3.7~e$^-$/keV), the recalculation of the previous estimate results in 38.5$\pm$6.9\% (red square in figure~\ref{img:EEE_SE_rec}) and lies closer to the NEST prediction.

\section{Conclusion}
\label{sec:concl}
A comprehensive calibration of the RED-100 detector, including several types of measurements, was performed. This paper outlines the data collection, processing, and subsequent analysis methodologies of the calibration data. 
The detector performances based on the \textit{in situ} calibration of single-photon and position-dependent PMT responses have been described.
One of the most significant and useful outcomes of the calibration is the calculation of the LRF shape using an iterative algorithm. 
LRFs can be applied to the position and energy reconstruction of all event types. 
Additionally, knowledge of the LRF shape is essential for detailed signal simulation.

The signals from single ionization electrons have been studied. 
Important parameters such as duration and single electron gain were obtained. Also, the stability of the detector parameters was monitored, and minor variations were corrected. 
The electron lifetime was measured and is described separately in~\cite{The_RED100_Experiment}. 
Light and charge calibrations were performed by means of gamma sources. 
Using the combined (S1+S2) energy scale, a good linearity of the response of the RED-100 detector was demonstrated. The EEE value was calculated with two approaches, giving similar results of 33.4$\pm$5.4\% and 32.8$\pm$2.8\% for the extraction electric field of 2.68±0.04 kV/cm. 
These results are in agreement with those of other experiments and the NEST predictions.

\acknowledgments
The RED-100 project was made possible thanks to administrative support from the State Atomic Energy Corporation Rosatom (ROSATOM) and the Rosenergoatom Joint-Stock Company and financial support from the JSC Science and Innovations (Scientific Division of the ROSATOM) under contract No.313/1679-D dated September 16, 2019 and from the Russian Science Foundation under contract No.22-12-00082 dated May 13, 2022. Authors express their gratitude for the National Research Nuclear University MEPhI (MEPhI Program Priority 2030), the National Research Center “Kurchatov Institute”, the Institute of Nuclear Physics named after G.I. Budker SB RAS, the Tomsk Polytechnic University (Development Program of Tomsk Polytechnic University No. Priority-2030-NIP/EB-004-0000-2022) for support in the development of technology of two-phase emission detectors. This work was funded by the Ministry of Science and Higher Education of the Russian Federation, Project "New Phenomena in Particle Physics and the Early Universe" FSWU-2023-0073. Also, the work was performed with the financial support provided by the Russian Ministry of Science and Higher Education, project “Neutrino detectors for remote monitoring of nuclear power plants and astrophysical installations”, No. FSWU-2022-0018. 

The authors are grateful to the staff of the Kalinin NPP for their continuous organizational and technical support during the RED-100 experiment, as well as the scientists from DANSS, $\nu$GeN, and iDREAM experiments at the Kalinin NPP, for assistance in organizing measurements.


\bibliographystyle{JHEP}
\bibliography{biblio.bib}






\end{document}